\documentclass[letterpaper,titlepage,11pt]{article}

\pdfoutput=1

\usepackage{amsmath}
\usepackage{amsfonts}
\usepackage{amssymb}
\usepackage{graphicx} 
\usepackage{subfigure}

\usepackage[
      colorlinks=true,
      linkcolor=blue,
      urlcolor=blue,
      filecolor=black,
      citecolor=red,
      pdfstartview=FitV,
      pdftitle={},
        pdfauthor={Roberto Emparan, Pau Figueras, Marina Martinez},
        pdfsubject={},
        pdfkeywords={},
        pdfpagemode=None,
        bookmarksopen=true
      ]{hyperref}

\def\clock{{\count0=\time
           \divide\count0 60
           \ifnum\count0<10 0\fi\the\count0
           \multiply\count0 -60 \advance\count0 \time
           :\ifnum\count0<10 0\fi \the\count0
         }}
\newcommand{\timestamp}{{\small\vbox{\hbox{\tt\jobname.tex}
\hbox{\the\day/\the\month/\the\year, \clock}}}}

\newcommand{\ie}{{\it i.e.,\,}}

\newcommand{\beq}{\begin{equation}}
\newcommand{\eeq}{\end{equation}}
\newcommand{\bea}{\begin{eqnarray}}
\newcommand{\eea}{\end{eqnarray}}
\newcommand{\beqa}{\begin{eqnarray}}
\newcommand{\eeqa}{\end{eqnarray}}

\numberwithin{equation}{section}

\begin{document}
\begin{titlepage}
\leftline{}
\vskip 2cm
\centerline{\LARGE \bf Bumpy black holes}
\vskip 1.2cm
\centerline{\bf Roberto Emparan$^{a,b}$, Pau Figueras$^{c}$, Marina Mart{\'\i}nez$^{b}$,}
\vskip 0.5cm
\centerline{\sl $^{a}$Instituci\'o Catalana de Recerca i Estudis
Avan\c cats (ICREA)}
\centerline{\sl Passeig Llu\'{\i}s Companys 23, E-08010 Barcelona, Spain}
\smallskip
\centerline{\sl $^{b}$Departament de F{\'\i}sica Fonamental, Institut de
Ci\`encies del Cosmos,}
\centerline{\sl  Universitat de
Barcelona, Mart\'{\i} i Franqu\`es 1, E-08028 Barcelona, Spain}
\smallskip
\centerline{\sl $^{c}$DAMTP, Centre for Mathematical Sciences,}
\centerline{\sl Wilberforce Road, Cambridge CB3 0WA, UK }
\smallskip
\vskip 0.5cm

\vskip 1.2cm
\centerline{\bf Abstract} \vskip 0.2cm 
\noindent 
We study six-dimensional rotating black holes with  \textit{bumpy} horizons: these are topologically spherical, but the sizes of symmetric cycles on the horizon vary non-monotonically with the polar angle. We construct them numerically for the first three bumpy families, and follow them in solution space until they approach critical solutions with localized singularities on the horizon. We find strong evidence of the conical structures that have been conjectured to mediate the transitions to black rings, to black Saturns, and to a novel class of bumpy black rings. For a different, recently identified class of bumpy black holes, we find evidence that this family ends in solutions with a localized singularity that exhibits apparently universal properties, and which does not seem to allow for transitions to any known class of black holes.

\end{titlepage}
\pagestyle{empty}
\small
\normalsize
\newpage
\pagestyle{plain}
\setcounter{page}{1}

\section{Introduction and main results}

In spite of the lack of effective solution-generating methods, the exploration of black hole solutions of the vacuum Einstein equations in $D\geq 6$ has made significant strides through the complementary use of approximate analytical methods and numerical calculations.
One line of study follows the observation that rapidly spinning Myers-Perry (MP) black holes in $D\geq 6$ \cite{Myers:1986un} approach black membranes, so they can be expected to admit, as black branes do, stationary deformations that ripple the horizon \cite{Emparan:2003sy}. Such \textit{bumpy black hole} solutions would naturally connect to black rings, black Saturns, and multi-ring solutions through topology-changing transitions in solution space \cite{Emparan:2007wm,Emparan:2010sx}. Evidence for this picture has been provided in \cite{Dias:2009iu,Dias:2010maa,Kleihaus:2012xh,Dias:2014cia,Kleihaus:2014pha}. In this article we confirm, refine, and extend aspects of it through a detailed numerical investigation of bumpy black holes in $D=6$. 

Bumpy black holes, like MP black holes (in the singly-spinning case that will be the focus of this article), have horizon topology $S^{D-2}$ with spatial symmetry group $U(1)\times SO(D-3)$. What distinguishes them from the ``smooth'' MP black holes is that the size of the $S^{D-4}$ symmetry orbits on their horizon varies in a non-monotonic fashion from the axis of rotation to the equator. 


The different families of bumpy black holes are conveniently identified by the way they branch off the MP family.
Refs.~\cite{Dias:2009iu,Dias:2010maa} identified linearized zero-mode perturbations of singly-rotating MP black holes, to which we can assign an `overtone' number $i=1,2,\dots$\footnote{We ignore the ``$i=0$'' zero mode \cite{Dias:2009iu} since it stays on the MP family and does not give rise to new branches of solutions.} that, for fixed mass, grows with the spin. Let us conventionally fix the sign of the zero-mode so that the $i$-th mode wavefunction at the axis of rotation has sign $(-1)^i$. By adding or subtracting the zero-mode perturbation to the MP black hole, we obtain two different branches, $(+)_i$ and $(-)_i$, of solutions emerging from the branching point. The evolution of the solutions along the $(+)$-branches was anticipated in \cite{Emparan:2007wm}: the horizon develops bumps that grow until a $S^{D-4}$ cycle pinches down to zero size, naturally suggesting a topology-changing transition to other black holes: black rings for $i=1$, black Saturns for $i=2$, and multi-ring configurations for $i>2$. The $(-)$-branches of solutions were only identified recently in \cite{Dias:2014cia}, and as we will see, they seem to terminate without any plausible connection to other black hole solutions.

\begin{figure}
  \centering
\subfigure[$(+)_1$-branch black hole at $j=1.13$, close to the transition to a black ring.]{\includegraphics[scale=0.7]{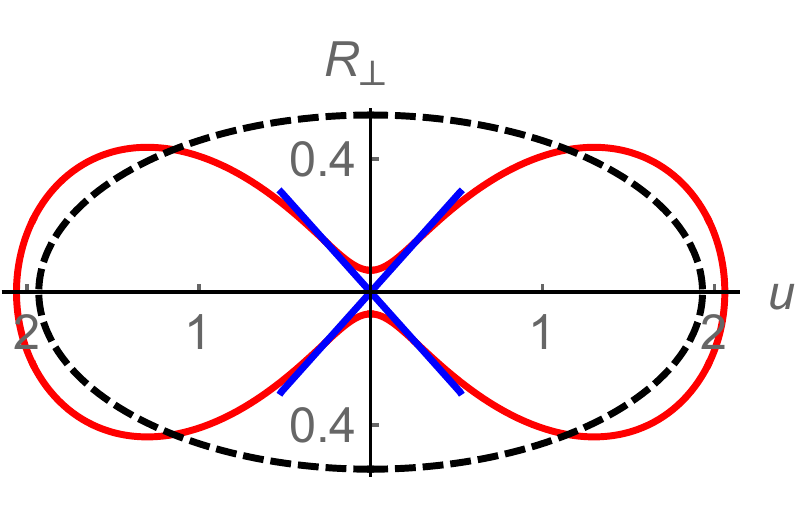}}\quad
\subfigure[$(+)_2$-branch black hole at $j=1.20$, close to the transition to a black Saturn. ]{\includegraphics[scale=0.7]{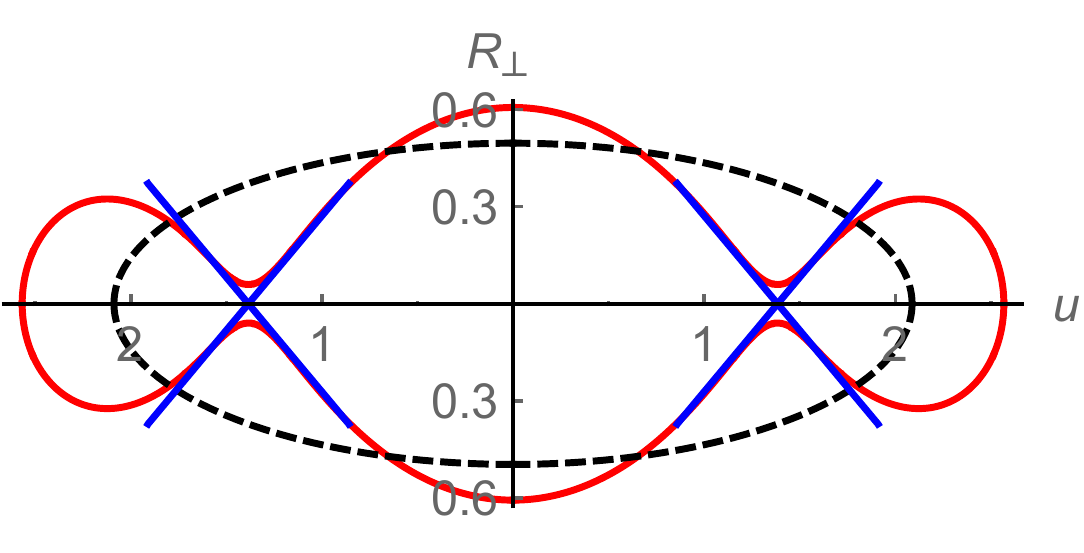}}\\
\subfigure[$(+)_3$-branch black hole at $j=1.55$, close to the transition to a bumpy black ring.]{\includegraphics[scale=0.7]{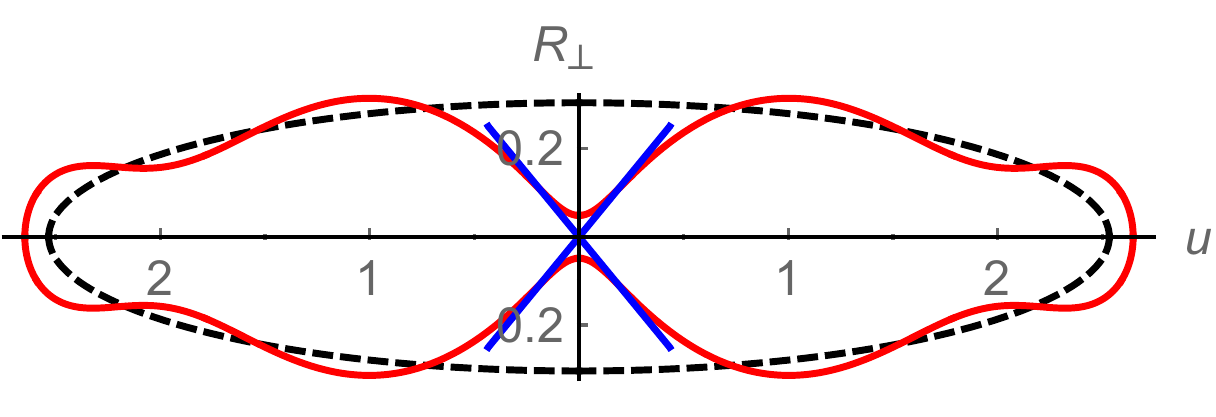}}
\caption{\small Embedding diagrams of bumpy black hole horizons of the $(+)_{1,2,3}$ branches (red curves), for the largest deformations we have obtained. The value of $R_\perp$ gives the size of the spheres $S^2$ transverse to the rotation plane. The vertical axis $u=0$ is the rotation axis, but $u$ does not measure the radius of the rotation circles. We superimpose  the embeddings of MP black holes with the same mass and spin (dashed black), and of the cones proposed for a local model of the critical singularity, eq.~\eqref{uzcones} (blue). The angular momentum $j$ is  normalized as in \eqref{dless}. In this and all subsequent plots, units are $GM=1$.}
\label{embdbrowncrit}
\end{figure}

\begin{figure}
  \centering
\includegraphics[scale=.6]{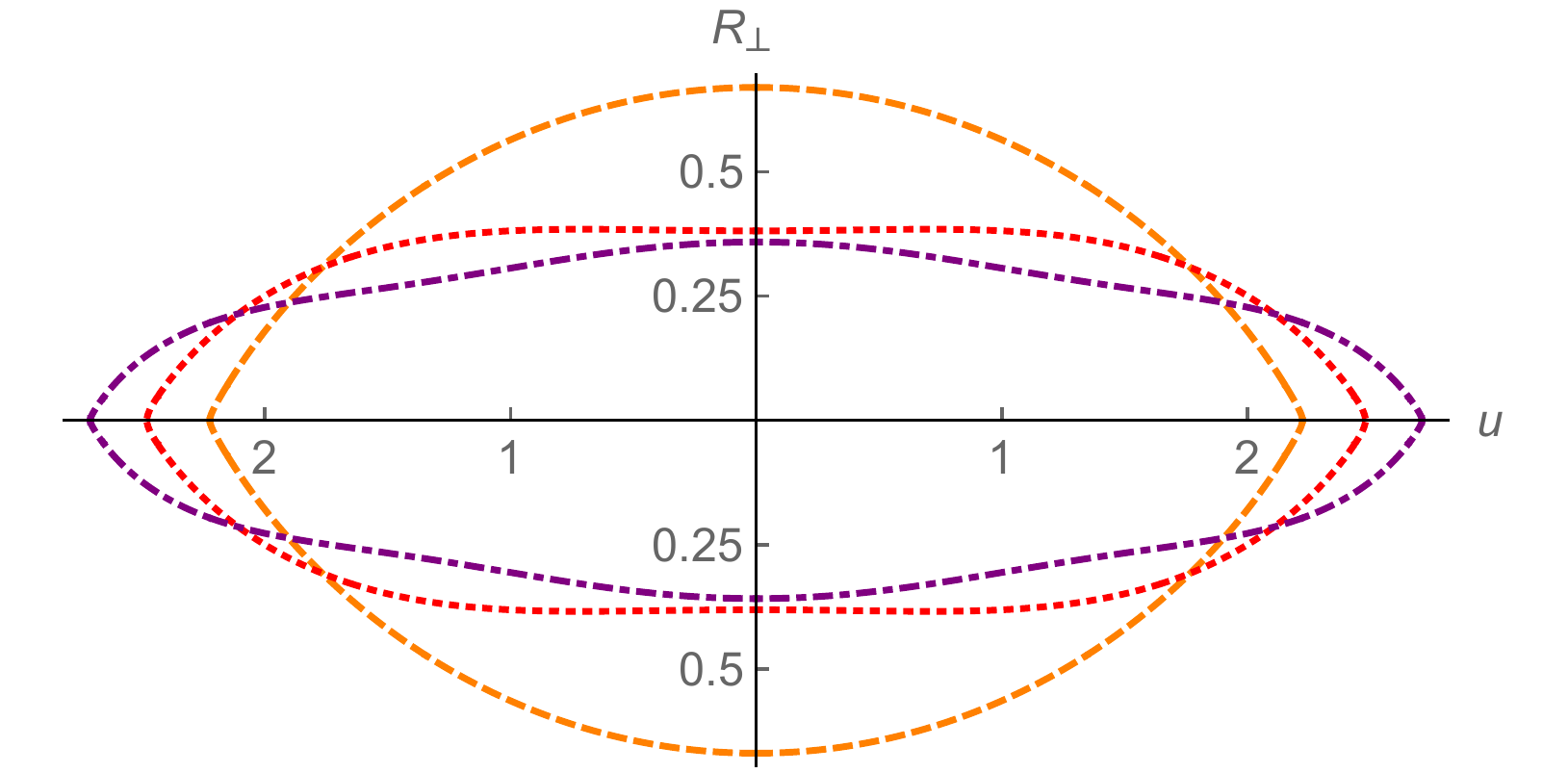}
\caption{\small Embedding diagrams for $(-)_{1,2,3}$ branch black holes (1: orange long-dash; 2: red short-dash; 3: purple dot-dash) at the largest deformations we have obtained. All branches exhibit the same singular conical shape near the equator (reflecting the rate at which the $S^2$ shrink there), with the same opening angle as in the critical $(+)_2$ solutions.}
\label{embdblackall}
\end{figure}

Ref.~\cite{Dias:2014cia} has studied the $(+)_1$ and $(-)_1$ branches in six and seven dimensions. Here we extend the analysis in the six-dimensional case to higher branches, $i=1,2,3$, while pushing both the $(+)$ and $(-)$ branches closer to their ends in phase space at singular solutions. We also perform a detailed investigation of the geometrical properties of these bumpy black holes. Our main conclusions, partly illustrated in figs.~\ref{embdbrowncrit} and \ref{embdblackall}, are:
\begin{enumerate}
\item The $(+)$ branches terminate at critical solutions with conifold-type singularities localized on the horizon of precisely the kind predicted in \cite{Emparan:2011ve} (following \cite{Kol:2002xz}).

\item The $(+)_3$ solutions pinch on the horizon at two places, on the rotation axis and off the axis, with the on-axis pinch growing deeper than the off-axis one. This strongly suggests that these solutions connect to a family of `bumpy black rings' not yet constructed. We expect that these rings eventually pinch off to connect to black di-rings.

\item The $(-)$ branches terminate at solutions with a curvature singularity localized at the equator of the horizon. The structure of the singularity appears to be locally the same for all $i$: the $S^2$ on the horizon shrinks to zero at the equator in a manner that resembles the cone that appears in the $(+)_2$ branch, while the length of the equatorial circle diverges. However, we are unable to provide an explicit local model for this singularity. We do not find any plausible extension of this branch to other singly-spinning black hole solutions.

\end{enumerate}

The conclusion in point 2 eliminates the possibility, considered as an alternative in \cite{Emparan:2007wm}, that the connection to black di-rings occurs through a phase of `bumpy black Saturns'. We give a simple argument to suggest that, as we move away of the MP solutions, the horizons in higher-$i$ branches pinch-off in succession from the rotation axis to the equator. Let us also remark that the divergent length of the equatorial circle mentioned in point 3 is not visible in fig.~\ref{embdblackall}, but will be made apparent in fig.~\ref{diagblack} below.

These results are explained in detail in sec.~\ref{sec:geom}, after having outlined in sec.~\ref{sec:constr} the construction of the solutions. In addition, in sec.~\ref{sec:phaseneg} we compute the thermodynamic properties of these solutions and draw phase diagrams. We also analyze the spectrum of the Lichnerowicz operator, and relate the number of negative eigenvalues to the thermodynamic stability of the solutions. The details of our numerics are relegated to appendix \ref{sec:numerics}. 

We remark that all these bumpy black holes are expected to be dynamically unstable; their importance lies in what they reveal about the possible geometries of black hole horizons in higher dimensions and the rich web of interconnections among them.

\section{Construction of the Solutions}\label{sec:constr}

In order to construct deformed rotating black holes in six dimensions we solve the Einstein-DeTurck equations $R^H_{ab}=0$ where
\begin{align}
R^H_{ab}=R_{ab}-\nabla_{(a}\xi_{b)}& &\text{ and }& &\xi^a=g^{bc}(\Gamma^a_{bc}-\bar{\Gamma}^a_{bc}).
\end{align}
$\Gamma$ is the usual Levi-Civita connection compatible with the spacetime metric $g$ and $\bar{\Gamma}$ is the Levi-Civita connection compatible with some reference metric $\bar{g}$ that satisfies the same boundary conditions as the spacetime metric $g$ but needs not be a solution to Einstein's equations. This is a standard method used in numerical General Relativity to find static and stationary solutions \cite{Headrick:2009pv,Figueras:2011va,Adam:2011dn}: the equations are manifestly elliptic and one can then use conventional numerical techniques for solving such partial differential equations.
For asymptotically flat (AdS or Kaluza-Klein) static metrics \cite{Figueras:2011va} proved that the solutions to the Einstein-DeTurck equations must in fact be Einstein. This result is yet to be extended to the stationary case, and hence, since we are interested in Einstein metrics, we must check that the DeTurck vector $\xi$ vanishes. For the solutions presented in this article we have checked that this is indeed the case, within our numerical accuracy.

The solutions we study are stationary and with only one of the two possible rotations turned on. Thus the rotation group $SO(5)$ is broken down to $U(1)\times SO(3)$, which act on the direction of rotation $\phi$ and on the spheres $S^2$ transverse to the rotation plane. The metric can then be written in the form
\begin{equation}
  \begin{split}
g_{ab}dx^adx^b=&-T(r,x)dt^2+P(r,x)(d\phi+W(r,x)dt)^2+S(r,x)d\Omega_{(2)}^2\\
&+A(r,x)dr^2+B(r,x)dx^2+2F(r,x)drdx
\end{split}
\label{ansatz}
\end{equation}
and we denote the reference metric as
\begin{equation}
  \begin{split}
\bar{g}_{ab}dx^adx^b=&-T_0(r,x)dt^2+P_0(r,x)(d\phi+W_0(r,x)dt)^2+S_0(r,x)d\Omega_{(2)}^2\\
&+A_0(r,x)dr^2+B_0(r,x)dx^2\,.
\end{split}
\label{refmetric}
\end{equation}
The compact radial direction $r\in [0,1)$  covers the region from the horizon at $r=0$, to infinity  at $r=1$. We seek solutions with horizons that are topologically $S^4$, so we choose sections at constant $t$ and $r$ to also be topological $S^4$'s. The size of the $\phi$-circles and of the symmetric $S^2$'s varies along the polar angular direction $x\in [0,1]$, with $x=0$ corresponding to the rotation axis (where $\phi$-circles shrink to zero) and $x=1$ to the equatorial plane (where $S^2$ spheres shrink to zero). The ``bumpiness'' of the horizon corresponds to non-monotonicity (of, say, the size of the $S^2$ on the horizon) along this polar direction. 

We write the metric functions as
\begin{equation}
  \begin{split}
  T(r,x)&=T_0(r,x)Q_{1}(r,x),\\ 
  W(r,x)&=W_0(r,z) Q_{2}(r,x),\\
  P(r,x)&=P_0(r,x)Q_{3}(r,x),\\
  S(r,x)&=S_0(r,x)Q_{4}(r,x),\\
  A(r,x)&=A_0(r,x)Q_{5}(r,x),\\
  B(r,x)&=B_0(r,x)Q_{6}(r,x),\\
  F(r,x)&=\frac{r x(1-x^2)}{(1-r^2)^3}Q_{7}(r,x)
\end{split}
\end{equation} 
and the reference metric is the MP metric with a small modification that enables us to control the temperature of the solutions.

In order to obtain the functions $(T_0,W_0,...\text{ etc.})$ for the reference metric we begin with the single-spin MP metric in standard Boyer-Lindquist-like coordinates (here $\bar{r}$, $\bar{\phi}$, $\bar{t}$ and $\theta$)
\begin{equation}
  \begin{split}
  ds^2=&-d\bar{t}^2+\frac{r_0^{3}}{\bar{r}\rho^2}\left(d\bar{t}+a\sin^2\theta d\bar{\phi}\right)^2+(\bar{r}^2+a^2)\sin^2\theta d\bar{\phi}^2\\
  &+\frac{\rho^2}{\Delta}d\bar{r}^2+\rho^2d\theta^2+\bar{r}^2\cos^2\theta d\Omega^2_{(2)}
  \end{split}
\end{equation}
where
\begin{equation}
 \rho^2=\bar{r}^2+a^2\cos^2\theta,\qquad \Delta=\bar{r}^2+a^2-\frac{r_0^3}{\bar{r}}
\end{equation}
and the $d\Omega_{(2)}^{2}$ is the line element of a 2-sphere. The horizon $(\bar{r}=r_+)$ is found by solving
\begin{equation}
  \Delta(r_+)=r_+^2+a^2-\frac{r_0^3}{r_+}=0,
\end{equation}
the mass and angular momentum are
\begin{equation}
M_{MP}=\frac{r_0^{3}\Omega_{(4)}}{4\pi G},\qquad J_{MP}=\frac{aM_{MP}}{2},
\end{equation}
 where $\Omega_{(4)}$ is the area of a unit 4-sphere, and the temperature and angular velocity are
\begin{equation}
    \mathcal{T}_{MP}=\frac{1}{4\pi}\left(\frac{2r_+^2}{r_0^3}+\frac{1}{r_+}\right),\qquad \Omega_H=\frac{a}{a^2+r_+^2}.
\end{equation}

We perform the changes of coordinates
\begin{equation}
  \bar{r}=\frac{r_+}{1-r^2},\qquad \cos\theta=1-x^2,\qquad\bar{\phi}=\phi+\Omega_H t,\qquad \bar{t}=t.
\end{equation}
The first two are made so that the ranges of the coordinates are $0<r,x<1$ and the third change, to co-rotating coordinates, is made because otherwise the $W_{0}$ function goes to zero too fast at infinity, which is inconvenient for numerical calculation. In co-rotating coordinates the function $W_0$ is 0 at the horizon and $\Omega_H$ asymptotically.  

The MP metric then takes the form
\begin{equation}
  \begin{split}
ds^2_{MP}=&-T_{MP}(r,x)dt^2+P_0(r,x)(d\phi+W_0(r,x)dt)^2\\
&+A_0(r,x)dr^2+B_0(r,x)dx^2+S_0d\Omega_{(2)}^2
\end{split}
\end{equation}
with
\begin{equation}
T_{MP}(r,x)=\frac{r^{2}\left(f(r)^{2}r_{0}^{3}+g(r)r_{+}^{3}\right)\left(f(x)^{2}\left(f(r)^{2}r_{0}^{3}+r^{2}g(r)r_{+}^{3}\right)+r^{2}x^{2}g(x)r_{+}^{3}\right)}{\left(f(r)^{2}r_{0}^{3}+r^{2}g(r)r_{+}^{3}\right)^{2}-r^{2}x^{2}f(r)^{2}g(x)\left(r_{0}^{3}-r_{+}^{3}\right)\left(f(r)^{2}r_{0}^{3}+g(r)r_{+}^{3}\right)},
\end{equation}
where
\begin{equation}
 f(r)=1-r^2,\qquad g(r)=2-r^2.
\end{equation}

We will find bumpy black hole solutions with given values of the temperature and angular velocity. It is convenient to specify these in terms of parameters of the reference metric. 
In order to control the temperature, we introduce a parameter $k$ in the reference metric
\begin{equation}
  T_0(r,x)=\frac{r^{2}\left(f(r)^{2}r_{0}^{3}k+g(r)r_{+}^{3}\right)\left(f(x)^{2}\left(f(r)^{2}r_{0}^{3}+r^{2}g(r)r_{+}^{3}\right)+r^{2}x^{2}g(x)r_{+}^{3}\right)}{\left(f(r)^{2}r_{0}^{3}+r^{2}g(r)r_{+}^{3}\right)^{2}-r^{2}x^{2}f(r)^{2}g(x)\left(r_{0}^{3}-r_{+}^{3}\right)\left(f(r)^{2}r_{0}^{3}+g(r)r_{+}^{3}\right)}
\end{equation}
so that the surface gravity $\kappa$ of the reference metric is given by
\begin{equation}
\kappa^2=\frac{T_0(r,x)}{r^2 A_0(r,x)}\Bigg|_{r\rightarrow 0}=\frac{1}{4r_+^2}\left(r_0^3k+2r_+^3\right)\left(r_0^3+2r_+^3\right).
\label{kappa}
\end{equation}
Obviously, whenever $k\neq 1$ the reference metric is not a solution of Einstein's equations 
but nonetheless it has a smooth horizon. However, $k$ allows us to move along the branches of solutions by varying it as a parameter in the reference metric. We will choose boundary conditions on the $Q$'s at the horizon in such a way that the surface gravity of the bumpy black holes is also given by (\ref{kappa}).
Note that by modifying $k$ not only the temperature  but also the mass and angular momentum of the solutions change. However, with the appropriate boundary conditions, $\Omega_H$ remains unchanged.

\subsection{Boundary conditions}
The conditions we impose on the $Q$'s at each of the boundaries of our domain in order to get regular solutions are

\textbf{Horizon $(r=0)$:}
The reference metric is already regular on the horizon. Since the spacetime metric is the reference metric multiplied by the $Q$'s, we ensure regularity on the horizon by imposing Neumann boundary conditions $\partial_r Q|_{r=0}=0$. In addition we impose $ Q_{1}(0,x)=Q_{5}(0,x)$, which fixes the surface gravity to the value (\ref{kappa}).

\textbf{Axis $(x=0)$:}
The reference metric is already regular on the axis of rotation so again we impose Neumann boundary conditions $\partial_x Q|_{x=0}=0$. The $\phi$ circle goes to zero at this boundary and in order to avoid a conical singularity we impose $Q_{3}(r,0)=Q_{6}(r,0)$.

\textbf{Equator $(x=1)$:}
The boundary conditions are again Neumann $\partial_x Q|_{x=1}=0$. Since here the radius of the $S^{2}$ shrinks to zero size, we impose $Q_{4}(r,1)=Q_{6}(r,1)$ to avoid a conical singularity.

\textbf{Infinity $(r=1)$:}
For asymptotically flat (AF) solutions, since the reference metric is already AF, we impose the Dirichlet boundary conditions 
\beq
Q_{i=1,\dots,6}(1,x)=1\,, \qquad Q_{7}(1,x)=0\,,
\eeq
so that the asymptotics are unchanged by the $Q$'s. Since we are in co-rotating coordinates the horizon angular velocity relative to infinity is given by the asymptotic value of $W(r,x)$. Then the condition $Q_{2}(1,x)=1$ ensures that $\Omega_H$ is given by the same expression as in the MP black hole.

\subsection{Physical magnitudes}
Given our choice of boundary conditions, the temperature and the angular velocity at the horizon are easily extracted in terms of quantities present in the reference metric, namely, $r_0$, $k$ and $r_+$, so that
\begin{equation}
  \mathcal{T}=\frac{1}{4\pi r_+}\sqrt{\left(r_0^3k+2r_+^3\right)\left(r_0^3+2r_+^3\right)},\qquad \Omega_H=\frac{\sqrt{r_+(r_0^3-r_+^3)}}{r_0^3}.
\end{equation}

Since we work with vacuum solutions, we can obtain the mass and angular momentum by evaluating their Komar integrals at the horizon,
\begin{equation}
  M=\frac{1}{12\pi G}\int_H*d\chi,\qquad J=\frac{-1}{16\pi G}\int_H*d\zeta,
\end{equation}
where $\chi$ is the 1-form dual to the asymptotic time-translation Killing vector $\partial_t-\Omega_H\partial_\phi$, and $\zeta$ is dual to the axial Killing vector $\partial_\phi$. In addition to these quantities we also compute the area of the horizon. In order to compare different solutions that have the same mass we use the dimensionless quantities 
\begin{equation}\label{dless}
\begin{split}
  &a_H^{D-3}=c_a\frac{\mathcal{A}^{D-3}_H}{(GM)^{D-2}},\qquad j^{D-3}=c_j\frac{J^{D-3}}{GM^{D-2}},\\
  &\omega_H=c_\omega\Omega_H(GM)^{1/(D-3)},\qquad t_H=c_t \mathcal{T}(GM)^{1/(D-3)},
  \end{split}
\end{equation}
with the numerical factors $c_a, c_j, c_\omega, c_t$ chosen as in \cite{Emparan:2007wm}.

Other geometric invariant quantities that are of interest for characterizing the solutions are the radii on the horizon ($r=0$) of the circles parallel to the plane of rotation, $R_\parallel(x)$, and of the spheres $S^2$ orthogonal to it, $R_{\perp}(x)$. They are given by
\beq
R_\parallel(x)=\sqrt{P(0,x)}\,,\qquad R_{\perp}(x)=\sqrt{S(0,x)}\,.
\eeq
We render these dimensionless by dividing them by $(GM)^{1/(D-3)}$ without any additional factors.

We will often use $j$ as the `control parameter' that changes along a branch of solutions. The bumpy branches extend over rather narrow ranges of $j$. They originate at bifurcation points in the MP family given respectively by
\beq\label{bifu}
(\pm)_{1,2,3}~\mathrm{beginning}: \quad j=1.20,\, 1.41,\, 1.57.
\eeq
The $(+)$-branches initially extend towards larger values of $j$, but then bend backwards towards decreasing $j$, which we have followed down to
\beq\label{plusend}
(+)_{1,2,3}~\mathrm{end}: \quad j=1.13,\, 1.20,\, 1.55.
\eeq
Along the $(-)$-branches, $j$ decreases away from the bifurcation, and the lowest values we have attained are
\beq\label{minend}
(-)_{1,2,3}~\mathrm{end}: \quad j=1.11,\,1.36,\,1.53.
\eeq

\section{Geometry of bumpy black holes}\label{sec:geom}

In this section we explore the geometry of the solutions, in particular of their horizons. Since we have pushed the new branches close to their endpoints in solution space, one purpose is to examine whether the critical solutions of $(+)$ branches have singularities modeled by Ricci-flat double-cone geometries that can mediate the transitions to black ring, black Saturn, and multi-ring solutions \cite{Emparan:2011ve}. Such structures in topology-changing transitions were first argued to be present in the context of Kaluza-Klein black holes \cite{Kol:2002xz} and have been extensively studied, see \cite{Wiseman:2002zc,Wiseman:2002ti,Kol:2003ja,Harmark:2003dg,Harmark:2003eg,Kol:2003if,Kudoh:2003ki,Elvang:2004iz,Kudoh:2004hs,Kleihaus:2006ee,Sorkin:2006wp,Kleihaus:2007cf,Headrick:2009pv,Figueras:2012xj}, and \cite{Horowitz:2012nnc} for a recent review of the subject.

Another aim is to get a better understanding of the solutions in the $(-)$ branches, in particular where and how these branches end.

The spatial horizon geometry, $r=0$, $t=\text{constant}$, is
\beq
ds^2_H=B(0,x)dx^2+R^2_\parallel(x)d\phi^2+R_{\perp}^2(x)d\Omega_{(2)}^2\,.
\eeq
In order to gain some intuitive understanding of these geometries, we perform two kinds of plots: embedding diagrams of sections of the horizon into Euclidean space, and plots of the invariant radii of the $S^1$ and $S^2$ symmetry cycles.

\paragraph{Embedding diagrams.}

Embeddings  in  Euclidean space provide useful and intuitive visualizations of the geometry. Here we use the same type of embeddings as ref.~\cite{Elvang:2006dd} presented for black rings.  On the spatial horizon geometry we choose a section $\phi=\mathrm{const.}$,
\begin{equation}
ds^2_{sec}=B(0,x)dx^2+R_{\perp}^2(x)d\Omega_{(2)}^2,
\label{horizonsec}
\end{equation}
and embed it in $\mathbf{E}_4$ 
\begin{equation}
ds_{\mathbf{E}_4}^2=du^2+d\rho^2+\rho^2 d\Omega_{(2)}^2
\label{E4}
\end{equation}
as a surface of the form
\begin{equation}
\rho=R_{\perp}(x),\qquad u=u(x)\,,
\end{equation}
so the induced geometry is
\begin{equation}
ds^2_{emb}=(R_{\perp}'(x)^2+u'(x)^2)dx^2+R_{\perp}^2(x)d\Omega_{(2)}^2\,.
\label{euclchange}
\end{equation}
The embedding is found by integrating
\beq
u(x)=\int_0^x d\bar x \sqrt{B(0,\bar x)-R_\perp'(\bar x)^2},
\label{sphereradius}
\eeq
which is possible since $B(0,x)\geq R_\perp'(x)^2$ for all our solutions. In our plots we present $R_\perp(x)$ versus $u(x)$. 

The coordinate $u$ does not have any invariant meaning as the radius of the rotational $S^1$'s, since this representation misses the information about $R_\parallel(x)$. For this, we employ a different type of plot.\footnote{Embedding the $(x,\phi)$ part of the horizon in this manner fails at large rotations, as in the case of the Kerr solution. A different kind of embedding is nevertheless possible \cite{Dias:2010maa}.}

\paragraph{Invariant-radii plots.}

These are plots of $R_\perp(x)$ versus $R_\parallel(x)$. Information about the length in the polar direction is lost now, which makes the horizon shapes in these plots look somewhat peculiar. 

\subsection{$(+)$-branch bumpy black holes}

\begin{figure}
  \centering
\subfigure[$(+)_1$ black hole at  $j=1.17$.]{\includegraphics[scale=0.7]{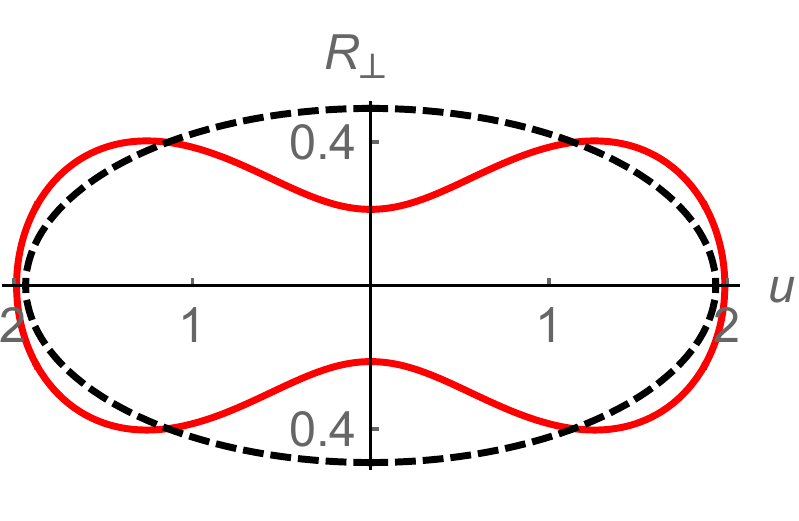}}\quad
\subfigure[$(+)_2$ black hole at  $j=1.36$.]{\includegraphics[scale=0.7]{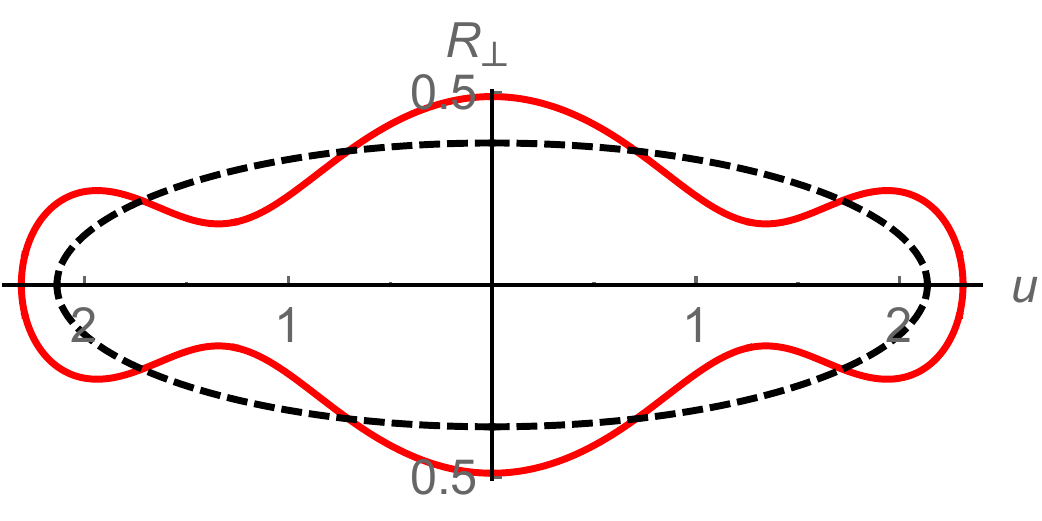}}\\
\subfigure[$(+)_3$ black hole at  $j=1.56$.]{\includegraphics[scale=0.7]{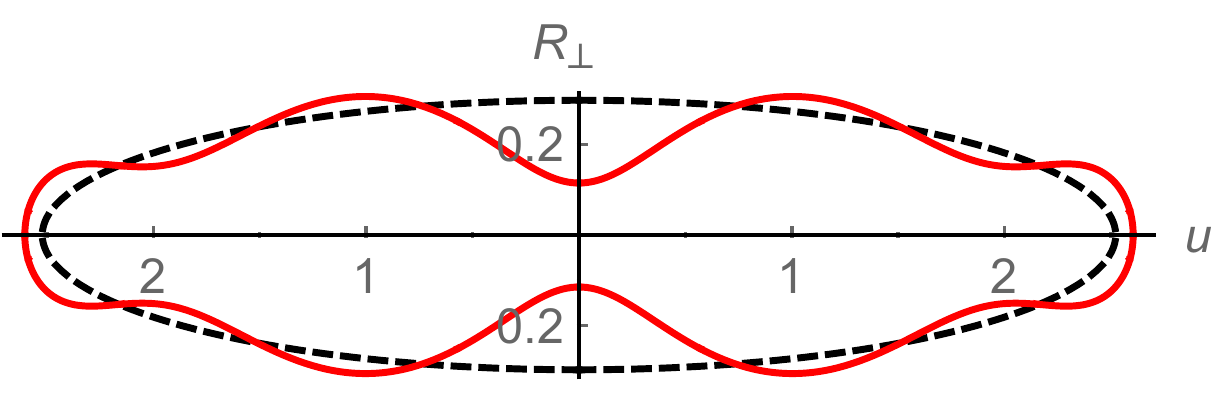}}
\caption{\small Isometric embeddings for representative black holes in the $(+)_{1,2,3}$ branches. $R_\perp$ is the radius of the $S^2$ orthogonal to the rotation plane and $u$ is a coordinate of Euclidean flat space, see (\ref{E4}) and (\ref{sphereradius}). The dashed black curve shows the embedding of a MP black hole of the same mass and angular momentum.}
\label{embdbrown}
\end{figure}

\begin{figure}
  \centering
\subfigure[$(+)_1$ black hole at  $j=1.17$.]{\includegraphics[scale=0.53]{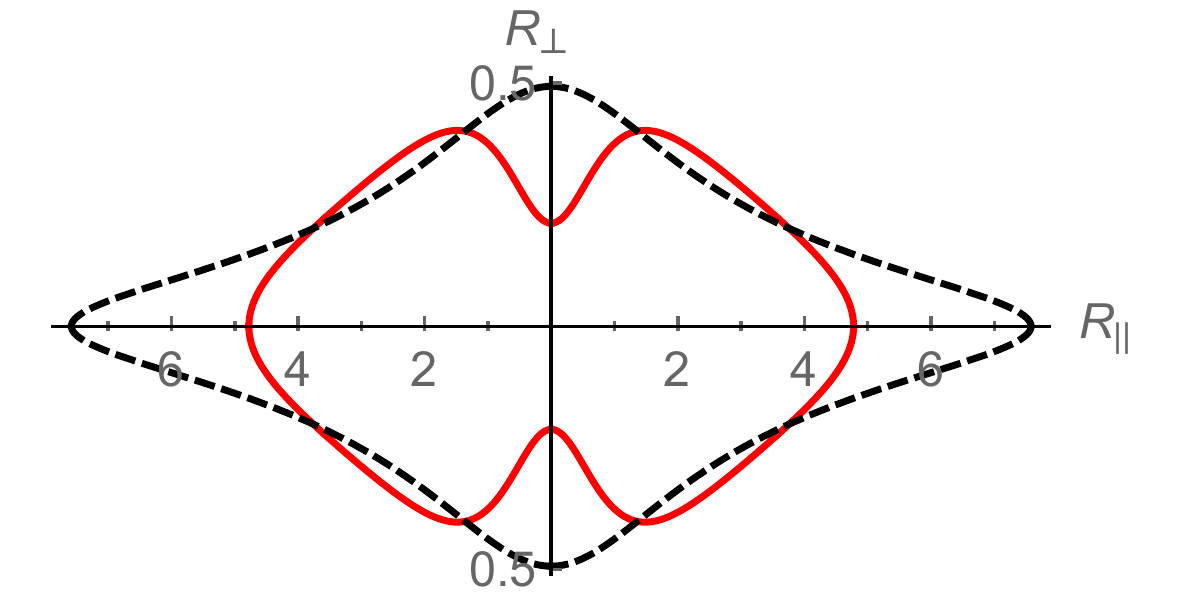}}\quad
\subfigure[$(+)_2$ black hole at  $j=1.36$.]{\includegraphics[scale=0.53]{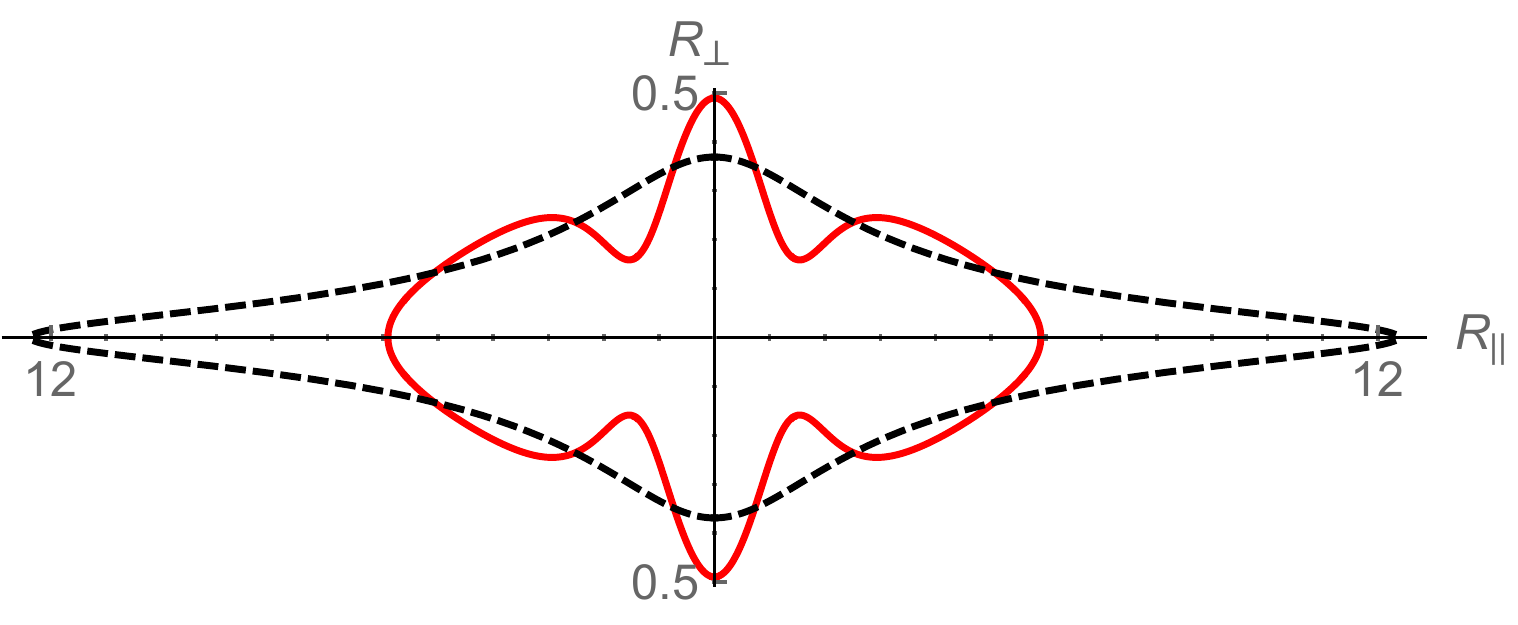}}\\
\subfigure[$(+)_3$ black hole at  $j=1.56$.]{\includegraphics[scale=0.53]{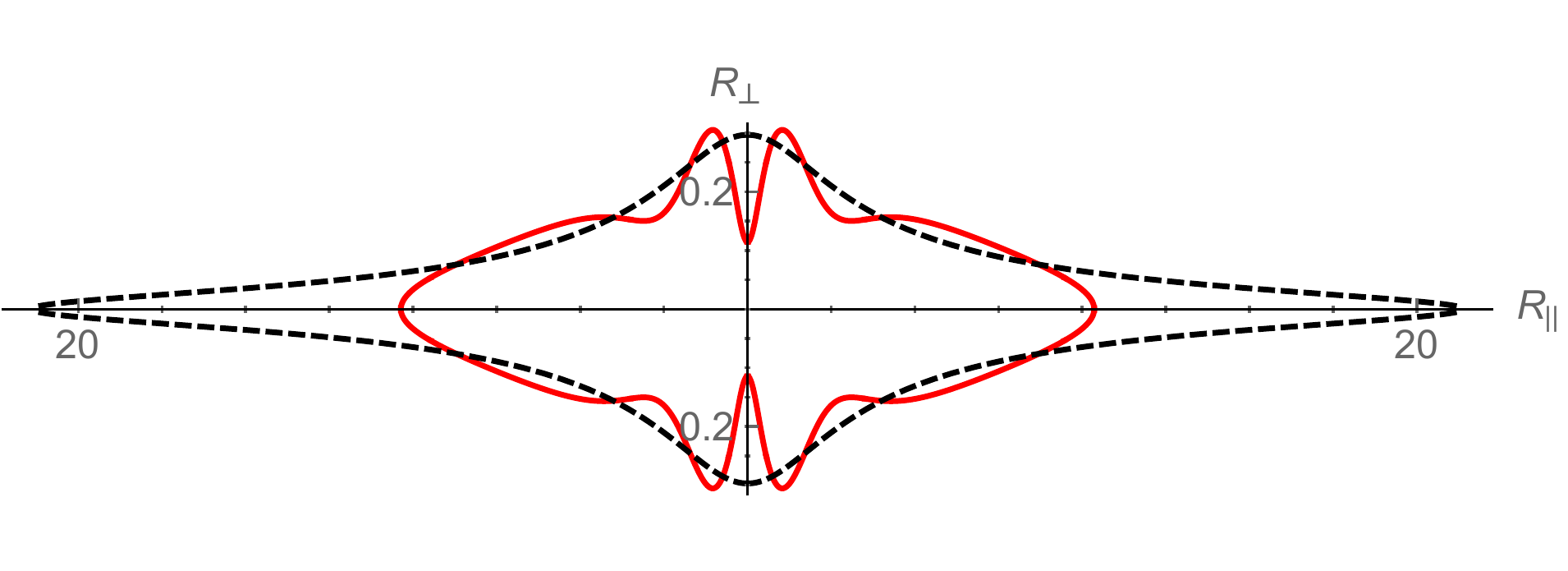}}
\caption{\small Invariant-radii plots for the same black holes as in fig.~\ref{embdbrown}. $R_\parallel$ is the radius of circles parallel to the rotation plane and $R_\perp$ is the radius of the orthogonal $S^2$. The black dashed curve shows a MP black hole of the same mass and angular momentum.}
\label{diagbrown}
\end{figure}

Representative solutions of these branches are depicted in embedding diagrams in fig.~\ref{embdbrown} and in invariant-radii diagrams in fig.~\ref{diagbrown}. Observe that, contrary to what may seem from the embedding diagrams, the radius $R_\parallel$ of the $S^1$ near the equator is larger in MP black holes than in the bumpy solutions with the same mass and angular momentum. 

Near the values \eqref{plusend} the solutions clearly approach configurations where a symmetric $S^2$ on the horizon pinches down to zero size, developing a singularity whose structure we analyze next.

\subsubsection{Critical cone geometries}

Depending on whether the singular pinch-off occurs along the rotation axis or on a circle away from the axis, the geometries are expected to be locally Lorentzian double-cones of the form
\begin{align}
ds^{2}_{\text{on-axis}}&=dz^2+\frac{2z^2}{D-2}\left(-\cos^{2}\chi dt^{2}+d\chi^{2}+\sin^{2}\chi d\phi^{2}+\frac{D-5}{2}d\Omega^2_{(D-4)}\right),\label{axiscone}\\
ds^{2}_{\text{off-axis}}&=dz^2+L^2d\phi^2+\frac{z^2}{D-3}\left(-\cos^{2}\chi dt^{2}+d\chi^{2}+(D-5)d\Omega^2_{{(D-4)}}\right),\label{circularcone}
\end{align}
with horizons at $\chi=\pi/2$, and where $L$ is the radius of the circle where the $S^2$ pinch to zero \cite{Emparan:2011ve}.

If we embed the section $t=\mathrm{const}$, $\chi=\pi/2$ of these geometries in Euclidean space as above, then it is easy to see that they are represented as the cones
\beq\label{uzcones}
\textrm{on-axis:}\quad u=\sqrt{\frac{3}{D-2}}z\,, \qquad \textrm{off-axis:}\quad u=\sqrt{\frac{2}{D-3}}z\,.
\eeq
We can superimpose these on the most deformed solutions we have obtained in these branches. Fig.~\ref{embdbrowncrit} shows excellent agreement with the prediction of \cite{Emparan:2011ve}.

\begin{figure}
  \centering
\subfigure[$(+)_1$-branch black hole at $j=1.13$, close to the transition to a black ring.]{\includegraphics[scale=0.53]{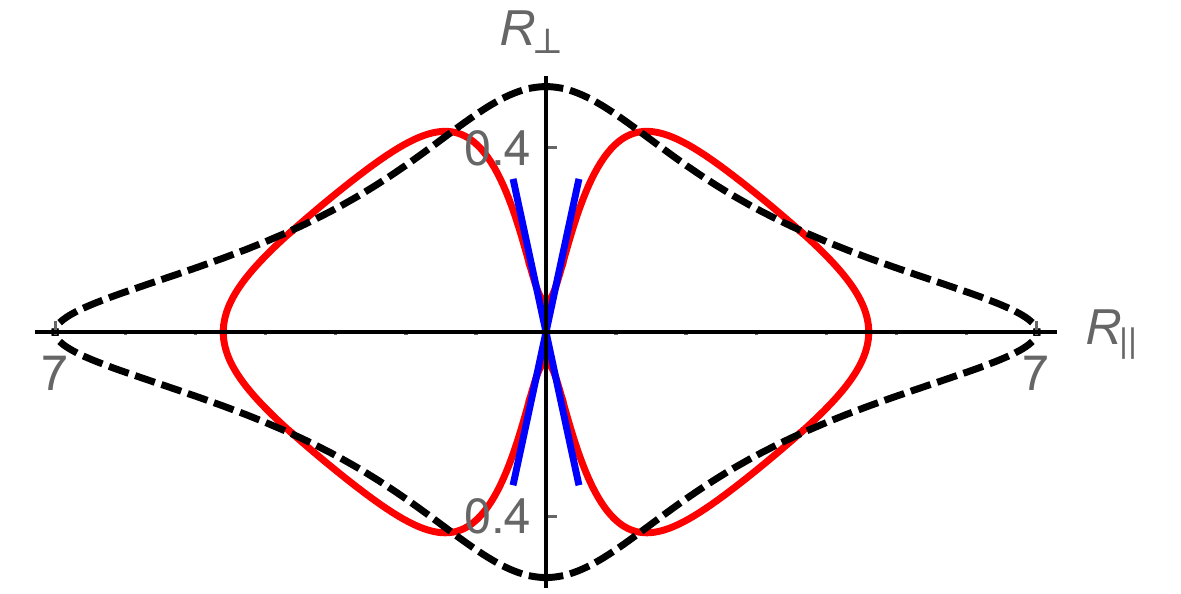}}\quad
\subfigure[$(+)_2$-branch black hole at $j=1.20$, close to the transition to a black Saturn. ]{\includegraphics[scale=0.53]{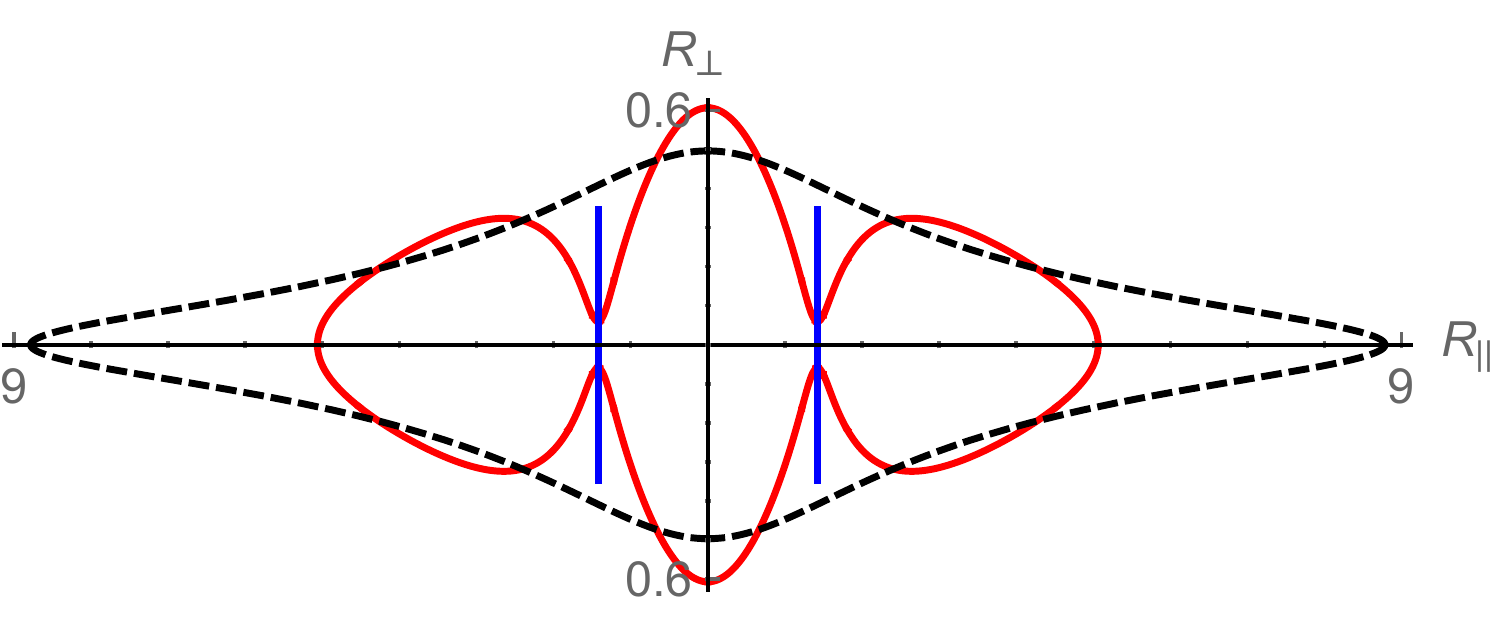}}\\
\subfigure[$(+)_3$-branch black hole at $j=1.55$, close to the transition to a bumpy black ring.]{\includegraphics[scale=0.5]{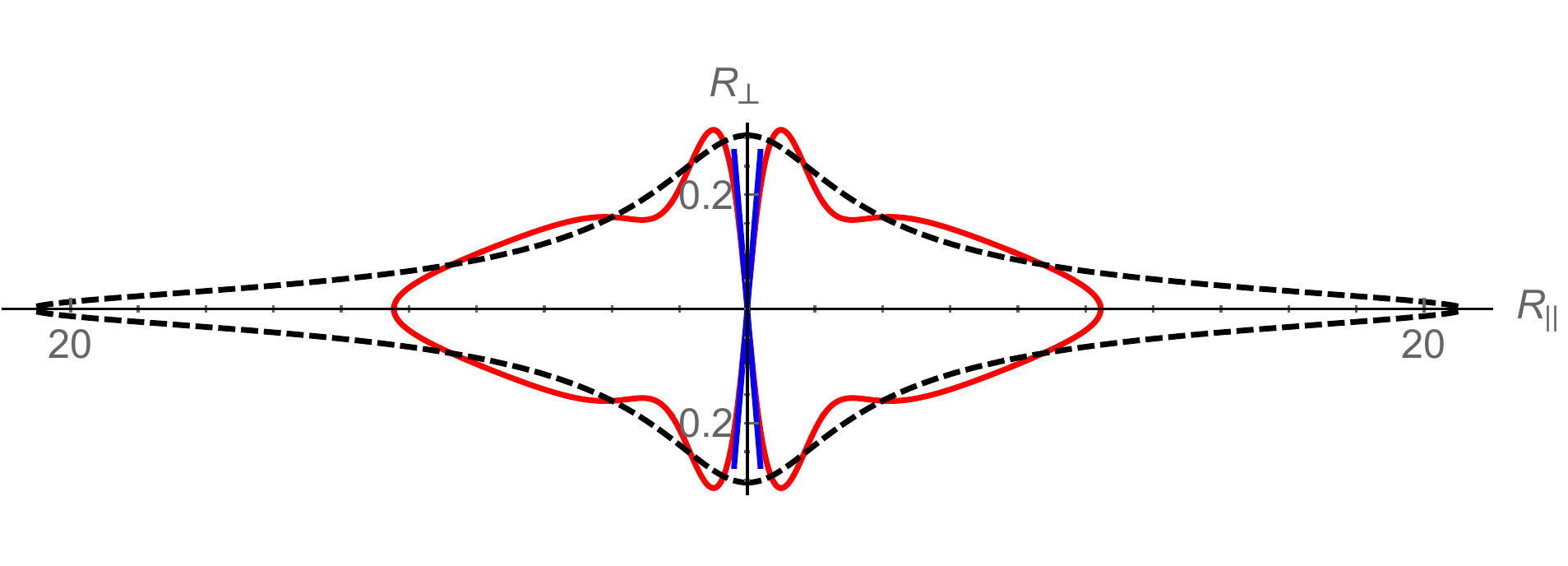}}
\caption{\small Invariant-radii diagrams of bumpy black hole horizons of the $(+)_{1,2,3}$ branches (red curves), for the largest deformations we have obtained. We superimpose the MP black holes with the same mass and spin (dashed black), and the plots for the conifolds  eq.~\eqref{radcones} (blue).}
\label{invbrowncrit}
\end{figure}

The invariant-radii plots probe complementary geometric aspects of the horizon. The geometries \eqref{axiscone}, \eqref{circularcone} have slopes
\beq\label{radcones}
\textrm{on-axis:}\quad \frac{dR_\perp}{dR_\parallel}=\sqrt{\frac{D-5}{2}}\,,\qquad \textrm{off-axis:}\quad \frac{dR_\perp}{dR_\parallel}\to\infty\,,
\eeq
which are also very well reproduced on-axis for $(+)_{1,3}$, see fig.~\ref{invbrowncrit}, but less well so off-axis for $(+)_{2}$, reflecting (maybe unsurprisingly) a remaining small dependence of $R_\parallel$ on the polar angle that would become negligible only much closer to the critical singularity.

We also compare the Kretschmann scalar 
\beq
K=R_{\mu\nu\rho\sigma}R^{\mu\nu\rho\sigma}
\eeq
of both geometries, following the study in \cite{Kol:2003ja} of the conical waist of inhomogeneous black strings. For the cones, $K$ depends only on the `polar' coordinate $z$ while for the black holes it depends not only on $x$ but also on $r$. In order to make the comparison we must specify a way to map points between the two geometries, \ie\ a function $z(r,x)$. This involves a certain arbitrariness, which we fix by equating the radius of the 2-sphere in both geometries. Then (in six dimensions)
\begin{equation}
\textrm{on-axis:}\quad S(r,x)=\frac{z^2}{4},\qquad 
\textrm{off-axis:}\quad S(r,x)=\frac{z^2}{3}
\eeq
and so
\beq
K_{\text{on-axis cone}}=\frac{72}{z^4}=\frac{9}{2S(r,x)^2}\,,
\qquad K_{\text{off-axis cone}}=\frac{48}{z^4}=\frac{16}{3S(r,x)^2}\,.
\end{equation}
These comparisons are dominated by how the size of the $S^2$ shrink close to the singularity, including away from the horizon, but they do not test the length of the equatorial $S^1$, to which $K$ is largely insensitive.  

We have computed the discrepancy between the Kretschmann scalars of both geometries, $\left|\frac{K_{\text{bh}}}{K_{\text{cone}}}-1\right|$, for the three branches and it is less than $10\%$ (often less than $5\%$) in the region near the singularity. Therefore, we conclude that the critical cones are locally a good description of the singular region.

\begin{figure}
  \centering
{\includegraphics[scale=0.7]{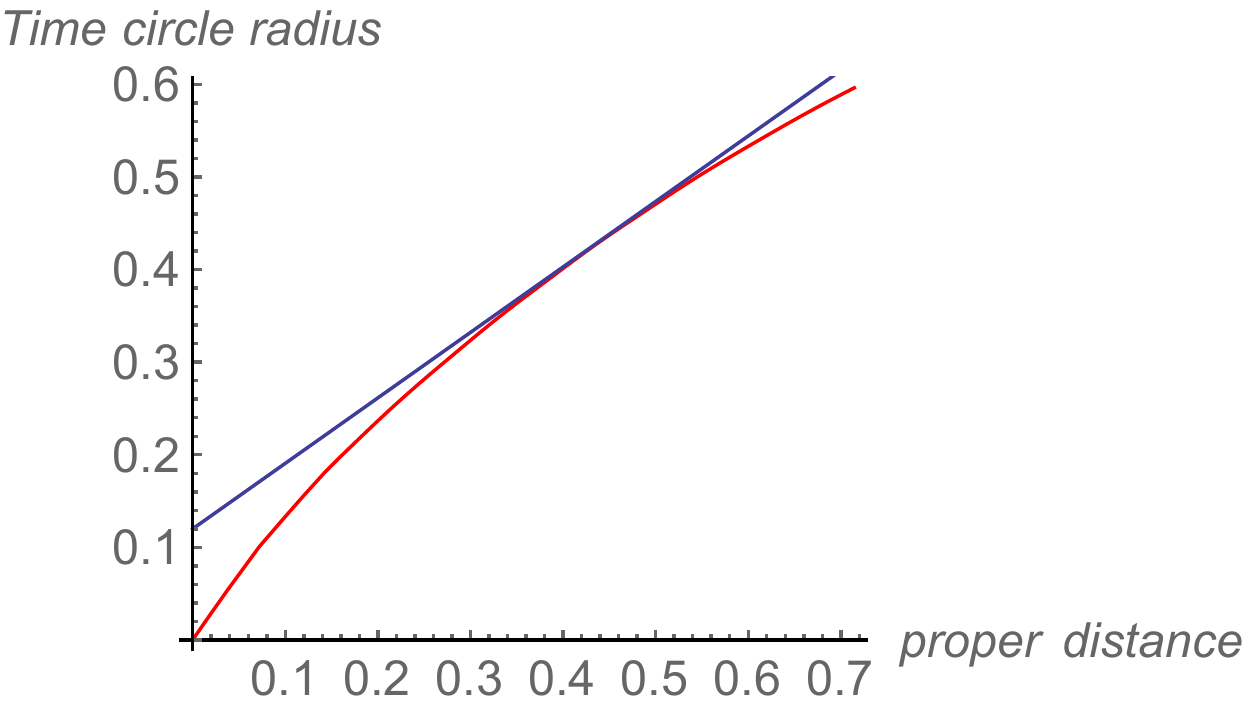}}
\caption{\small Circular-radius of the Euclidean time circle as a function of the proper distance to the horizon along the rotation axis, in the $(+)_1$ black hole at $j=1.13$. The slope matches well that of the cone geometry (blue) as the black hole is approached, although not very close to the horizon where the singular cone is smoothed in our solution.}
\label{timecones}
\end{figure}

Finally, we have also checked the appearance of a conical structure in the Euclidean time direction. Fig.~\ref{timecones} shows the rate at which the Euclidean time circle shrinks along the axis of rotation in our nearest-to-critical $(+)_1$ solution.\footnote{Close to the horizon, and in corotating coordinates, the geometry is very approximately static and one can sensibly talk about the Euclidean time circle.} The slope in this curve fits well the slope of the conical solution over a range of distances close to the black hole. It departs from it very near the horizon, as it must since the cone is smoothed in our solution\footnote{In fact, very close to the horizon the slope in this plot must become the same as the surface gravity.}.

\subsubsection{$(+)_3$: transition to bumpy black rings}

It was naturally conjectured in ref.~\cite{Emparan:2007wm} that black holes along the $(+)_{1,2}$ branches would pinch to zero and transition to black ring and black Saturn phases, respectively.  
However, higher branches $(+)_{i\geq 2}$ have multiple pinches and it was less clear what their fate could be. 
If pinch-down occurred first on a circle off-axis, then the branch $(+)_3$ would transition to a black Saturn configuration with a bumpy central black hole. However, the deformation of $(+)_3$ black holes is expected to be larger on-axis than off-axis. The reason is that in the black membrane limit of the MP black holes, and for small, linearized perturbations, the axisymmetric Gregory-Laflamme-type perturbation takes the form \cite{Emparan:2003sy}
\begin{equation}
\delta g_{\mu\nu}\sim J_{0}(x)h_{\mu\nu}(r),
\end{equation}
where $x$ is the distance from the rotation axis in directions parallel to the horizon, and hence plays the role of the polar angle. The Bessel function $J_0(x)$ yields larger deformations close to the axis of rotation at $x=0$, and decays away from it. Figs.~\ref{embdbrowncrit} and \ref{embdbrown} show that this behavior persists when the deformations are not small. 

This evolution of the $(+)_3$ branch has a natural end at a topology-changing transition to a branch of \textit{bumpy black rings}, of horizon topology $S^1\times S^{D-3}$, with a deformed $S^{D-3}$. These have not been constructed yet, and our arguments are the first clear indication of their existence. It is also natural to expect that the bumpy black ring branch will connect, at its other end, to black di-rings. Indeed it seems implausible that they smooth out their deformations and connect to the known (smooth) black ring solutions, since these two branches are very far apart in solution space (see fig.~\ref{thermoplots} below).

The argument also suggests that the same behavior occurs in higher branches, with pinches being larger closer to the axis, and pinching-down sequentially at increasing values of the polar angle $x$. The conifold-type transition then connects them to new families of multiply-bumpy black rings, which eventually, through several transitions, connect to multi-ring configurations.

\subsection{$(-)$-branch bumpy black holes}
\begin{figure}
  \centering
\subfigure[$(-)_1$-branch black hole with $j=1.11$.]{\includegraphics[scale=0.63]{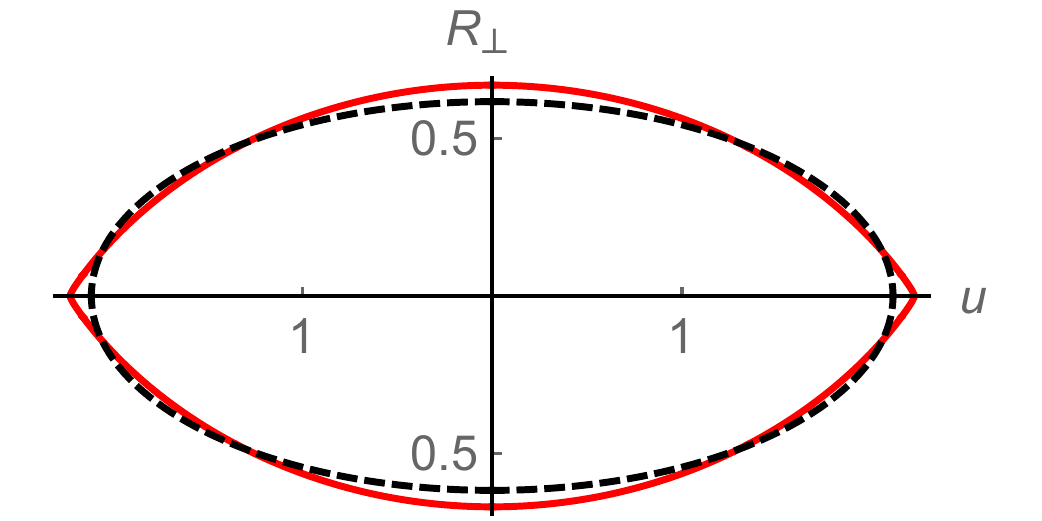}}\quad
\subfigure[$(-)_2$-branch black hole with  $j=1.36$.]{\includegraphics[scale=0.63]{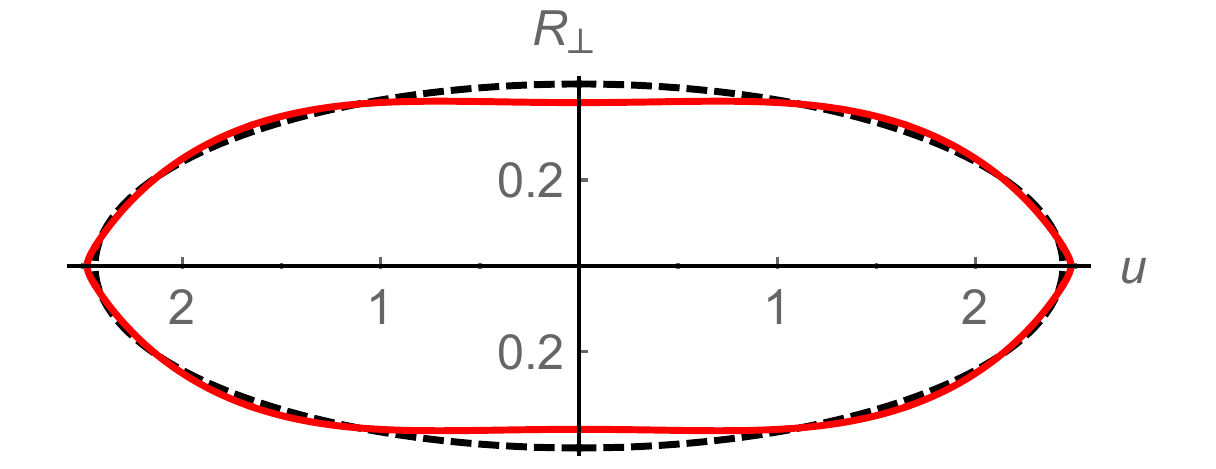}}\\
\subfigure[$(-)_3$-branch black hole with  $j=1.53$.]{\includegraphics[scale=0.63]{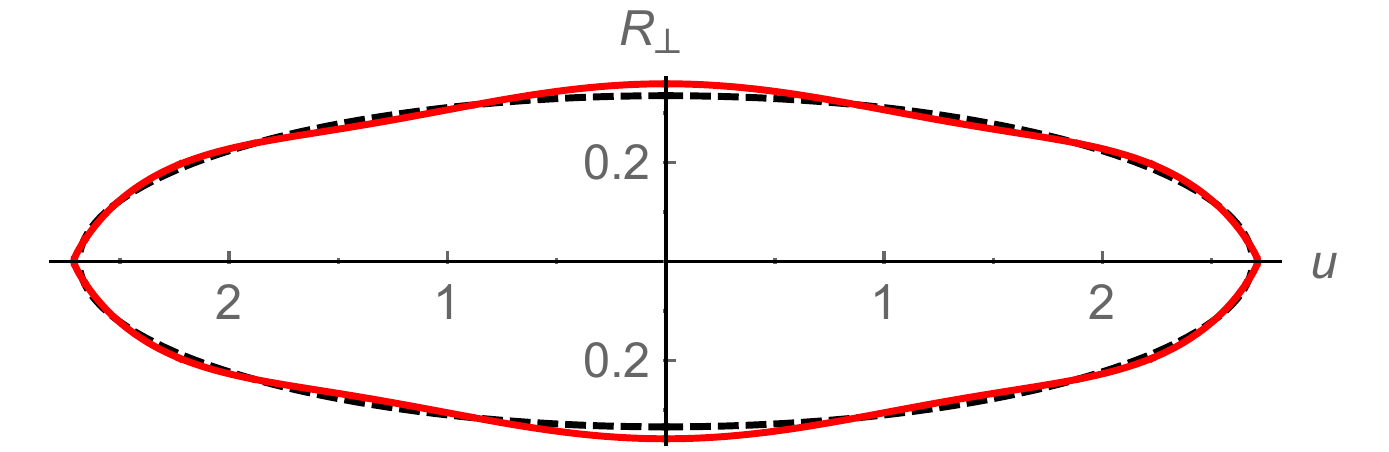}}
\caption{\small Isometric embeddings for bumpy black holes near the termination of the $(-)$ branches. $R_\perp$ is the radius of the transverse $S^2$ and $u$ is a coordinate of Euclidean flat space. The black dashed curve shows the embedding of a MP black hole of the same mass and angular momentum. The conical shapes at the equator have the same opening angle for the three branches, $u=\sqrt{2/3}\,z$.}
\label{embdblack}
\end{figure}

\begin{figure}
  \centering
\subfigure[$(-)_1$-branch black hole with $j=1.11$.]{\includegraphics[scale=0.5]{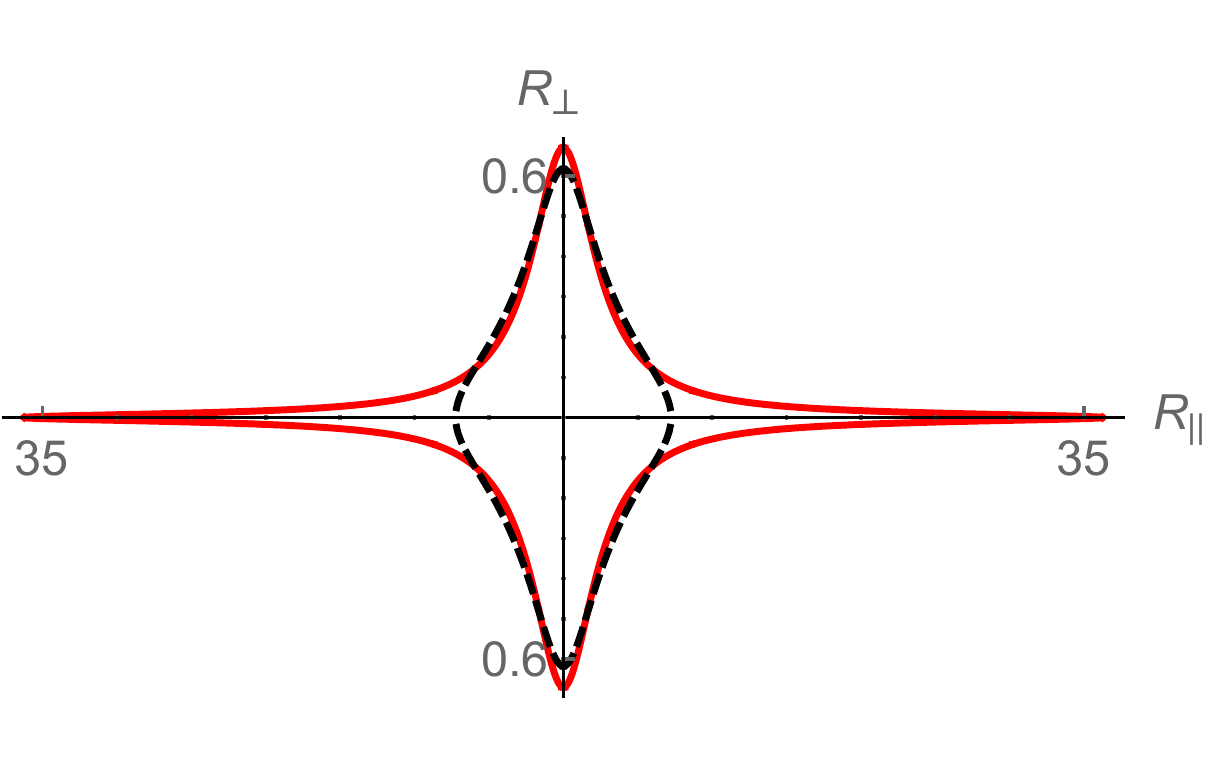}}\quad
\subfigure[$(-)_2$-branch black hole with  $j=1.36$.]{\includegraphics[scale=0.5]{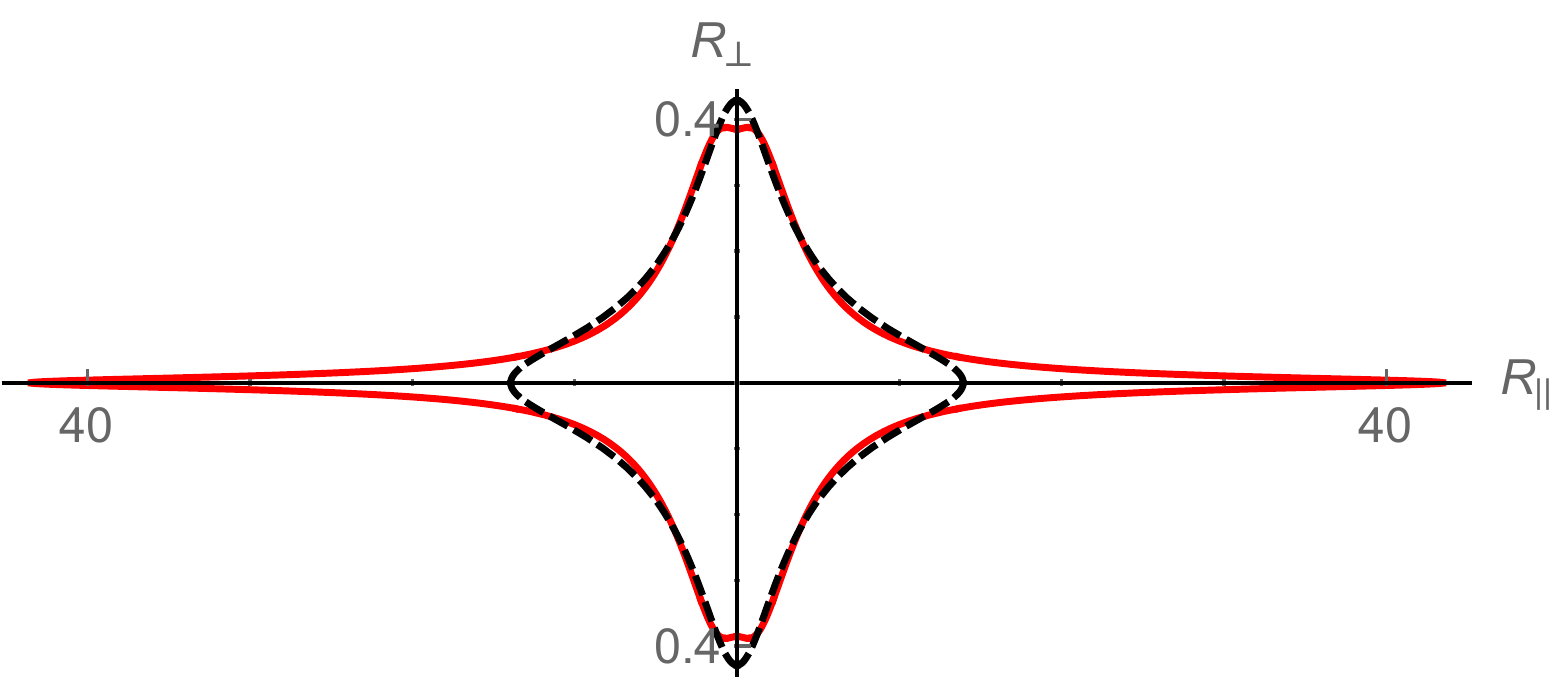}}\\
\subfigure[$(-)_3$-branch black hole with  $j=1.53$.]{\includegraphics[scale=0.5]{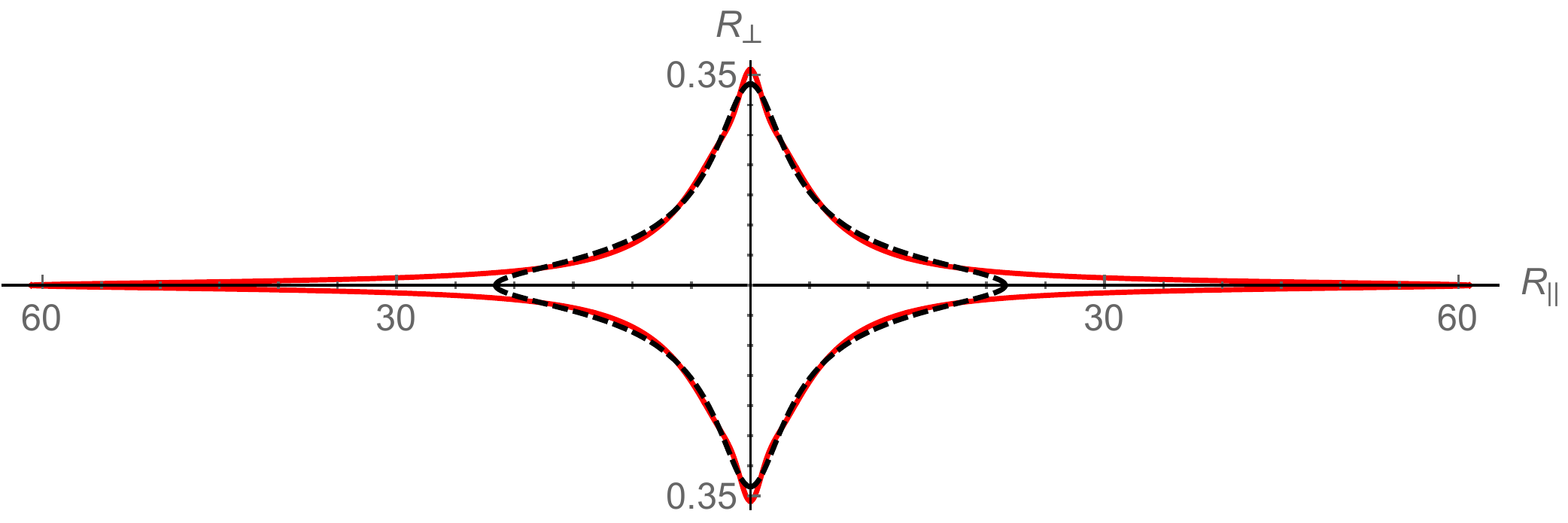}}
\caption{\small Invariant-radii plots for the same black holes as in fig.~\ref{embdblack}. $R_\parallel$ is the radius of circles parallel to the rotation plane and $R_\perp$ is the radius of the orthogonal  $S^2$. The black dashed curve shows a MP black hole of the same mass and angular momentum.}
\label{diagblack}
\end{figure}

Figs.~\ref{embdblack} and \ref{diagblack} show the previous two types of graphics for the horizon geometry of these black holes at their largest deformation, \eqref{minend} (these were also shown in fig.~\ref{embdblackall}). 

From fig.~\ref{diagblack} we see that these horizons spread in the rotation plane more than in the MP black holes of the same mass and spin. This could be anticipated near the bifurcation point, where the deformation is controlled by a zero-mode with $i+1$ nodes: since the $(-)_i$ zero mode wavefunctions have sign $(-1)^{i+1}$ at the rotation axis, then the wavefunction at the equator must always be positive, \ie\ the bumpy black hole bulges out.\footnote{The same argument implies that $(+)$-branch black holes bulge out less at the equator than MP black holes of the same mass and spin, see fig.~\ref{diagbrown}.}
At least for the $i=1$ solutions, we can also understand this in more physical terms: close to the branching point both solutions have the same mass, angular momentum and angular velocity. If the MP black hole is perturbed in such a way that some of its mass is concentrated closer to the axis of rotation, then in order to maintain the angular momentum constant (with the same angular velocity) some mass must also be moved farther along the rotation plane, preferrably around the equator.  

Further along the branch the horizons stretch a lot on the rotation plane, see fig.~\ref{diagblack}, and get highly pancaked, $R_\perp\ll R_\parallel$. Nevertheless, in contrast to ultraspinning MP black holes, they do not seem to approach black membranes in the limit, and in particular (as we will see in sec.~\ref{sec:phaseneg}) they do not develop the Gregory-Laflamme zero modes of the Lichnerowicz operator characteristic of black membranes that would signal the appearance of new branches of solutions \cite{Gregory:1987nb}.

\begin{figure}
  \centering
\includegraphics[scale=0.7]{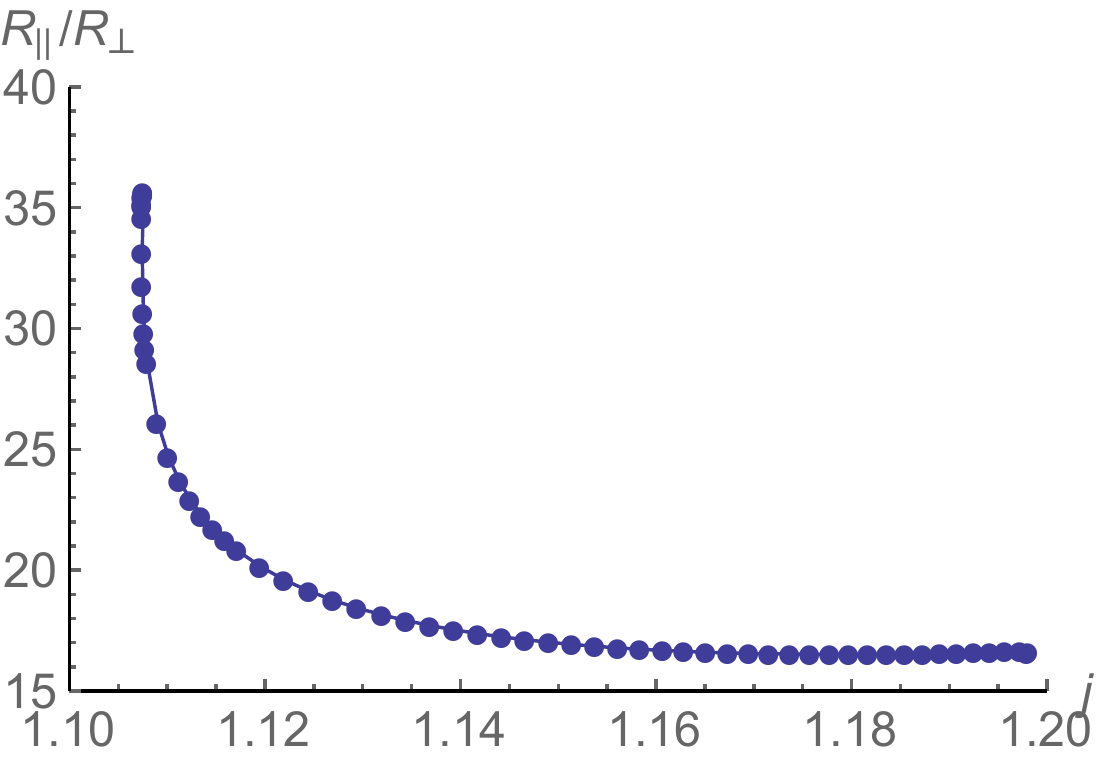}
\caption{\small $R_\parallel/R_\perp$ for $(-)_1$ bumpy black holes as a function of $j$. Note that $j$ decreases as the solutions get farther from the MP bifurcation point. Close to the limiting value $j\approx 1.11$ the equatorial radius $R_\parallel$ appears to diverge, both for fixed $R_\perp$ and for fixed mass. The $(-)_{2,3}$ branches show similar behavior.}
\label{RRratio}
\end{figure}

Fig.~\ref{RRratio} strongly suggests that the length of the equatorial circle diverges in the limiting solutions --- even though the radial distance to the equator remains finite. This behavior is known to occur for the extremal limit of the five-dimensional MP black hole, although in the latter case the extremal solution has zero temperature and area, whereas these remain finite in the critical $(-)$ solutions.

The $S^2$'s on the equator shrink to zero in the limiting solutions in a singular way, causing the Kretschmann scalar to diverge. The effect seems to be the same in all three branches, being
well reproduced on sections of constant $t$ and $\phi$ on the horizon (such as are captured in fig.~\ref{embdblack}, and by the Kretschmann scalar) by the geometry
\beq\label{s2cone}
ds^2=dz^2+\frac{z^2}{3}d\Omega_{(2)},
\eeq
which is also present in off-axis cones \eqref{circularcone}.

This suggests that the local structure of the singularity at the equator in these solutions may be universal for all $(-)$ branches: the $S^2$ shrink to zero along the horizon like in \eqref{s2cone}, while the length of the equatorial $S^1$ diverge.

Although we do not have a local model for the full singularity, it is not one of the conical geometries that effect a transition to another branch of black holes. In fact it seems unlikely that the singularity is a Ricci-flat scaling geometry. In view of this, and in the absence of a plausible candidate for a merger transition, we are led to conjecture that the $(-)$ branches of black holes terminate in phase space without continuing into any other singly-spinning stationary black hole solutions.

\section{Phase diagrams, thermodynamic stability, and negative modes}
\label{sec:phaseneg}

\begin{figure}
  \centering
\subfigure[Area $a_{H}$ vs.\ spin $j$]{\includegraphics[scale=0.5]{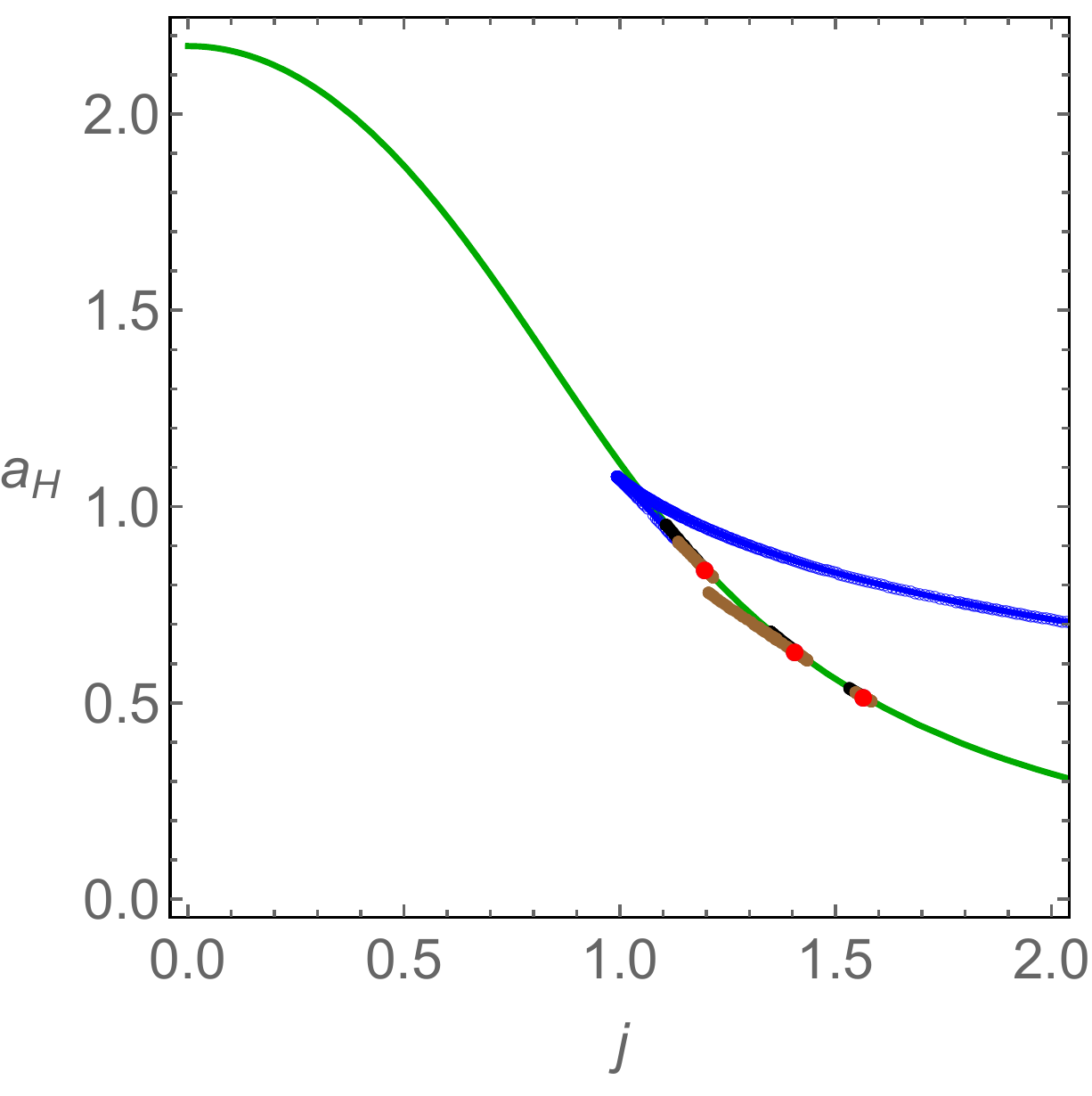}}\qquad
\subfigure[Zoom of (a)]{\includegraphics[scale=0.59]{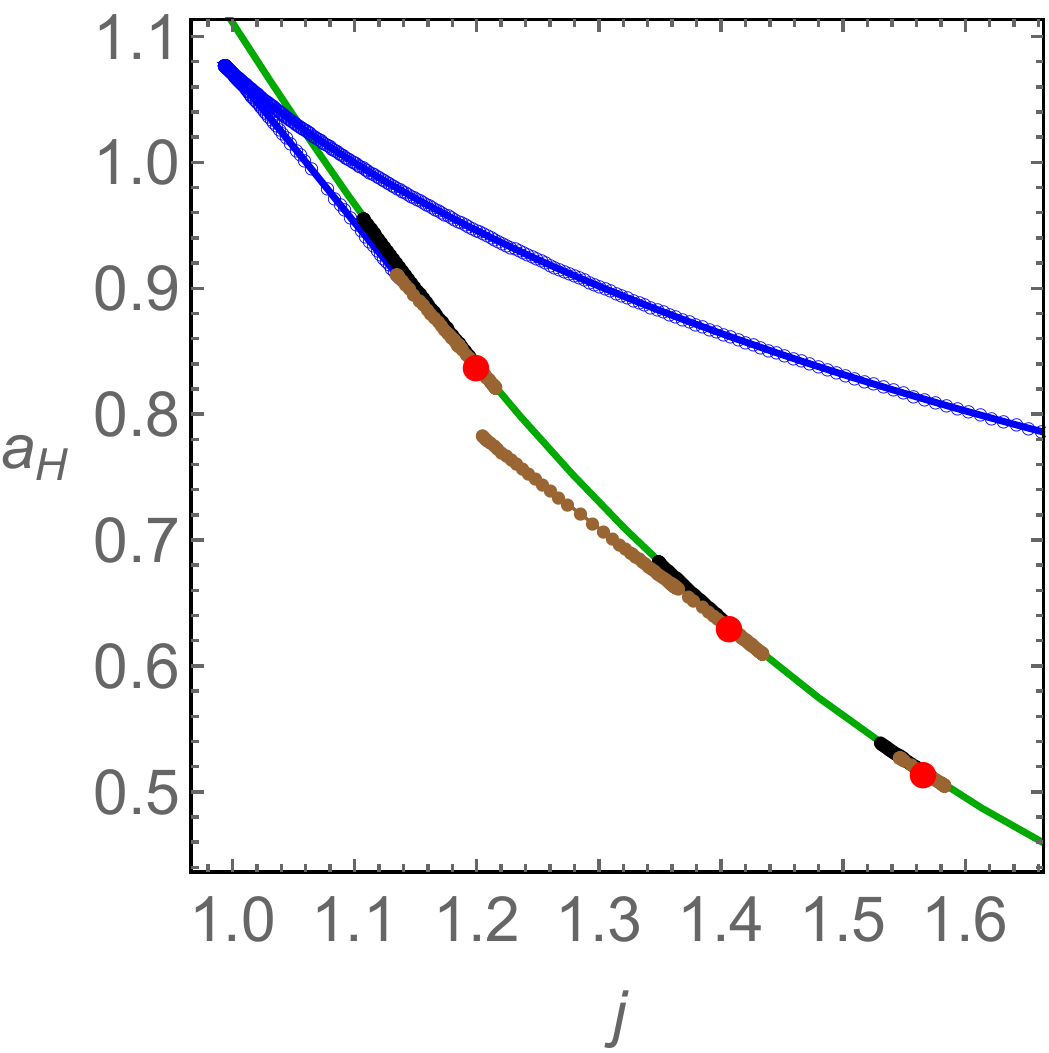}}\\
\subfigure[Temperature $t_H$ vs.\ spin $j$]{\includegraphics[scale=0.5]{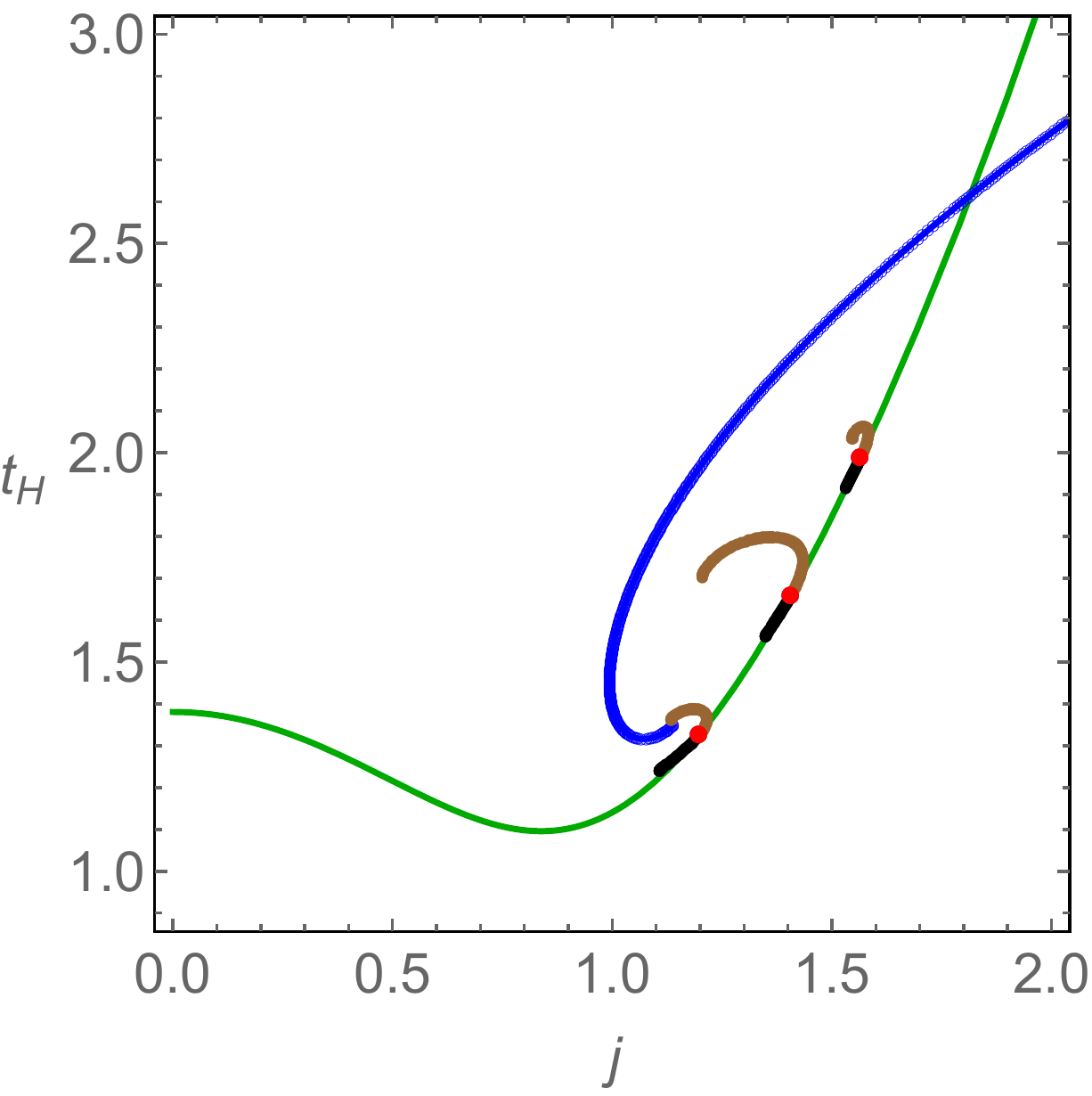}}\qquad
\subfigure[Zoom of (c)]{\includegraphics[scale=0.57]{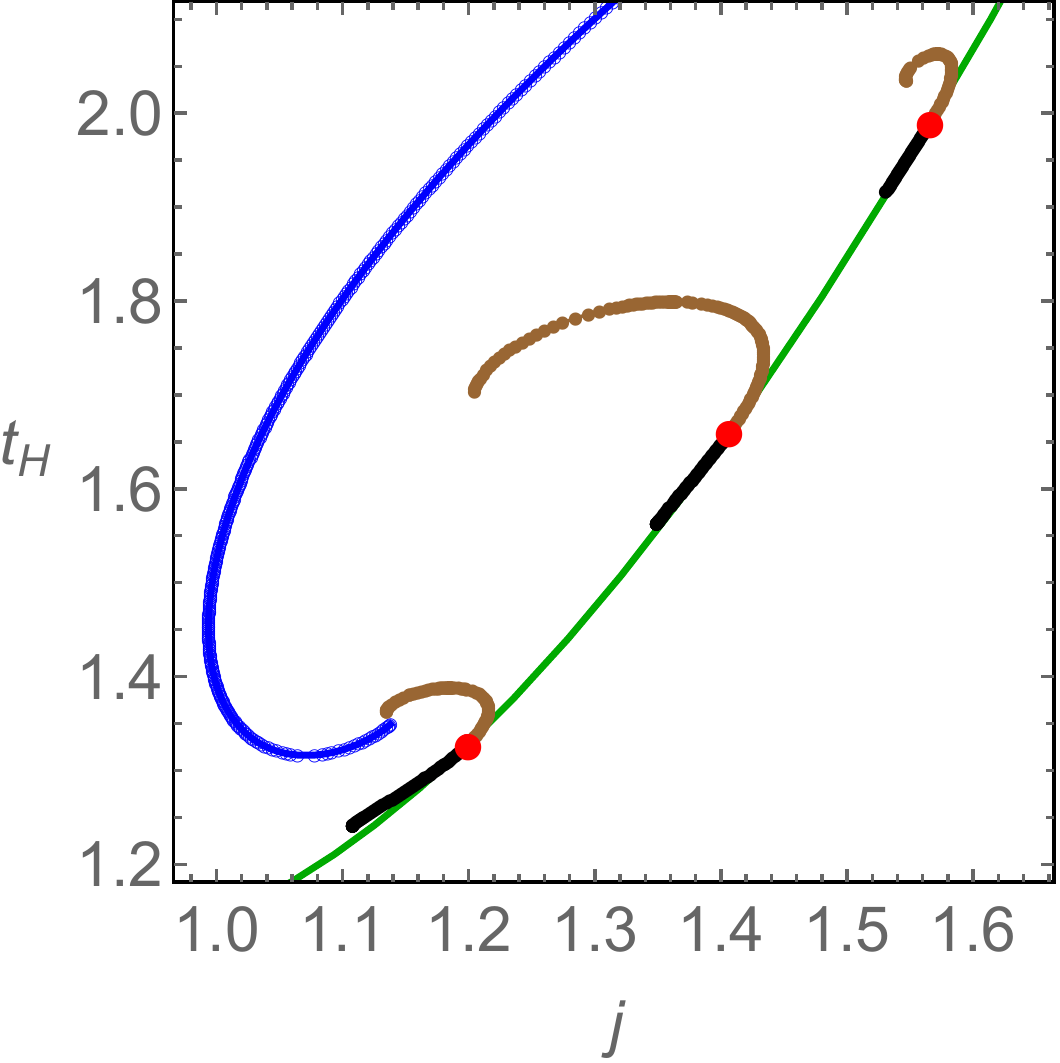}}\\
\subfigure[Angular velocity $\omega_H$ vs.\ spin $j$]{\includegraphics[scale=0.5]{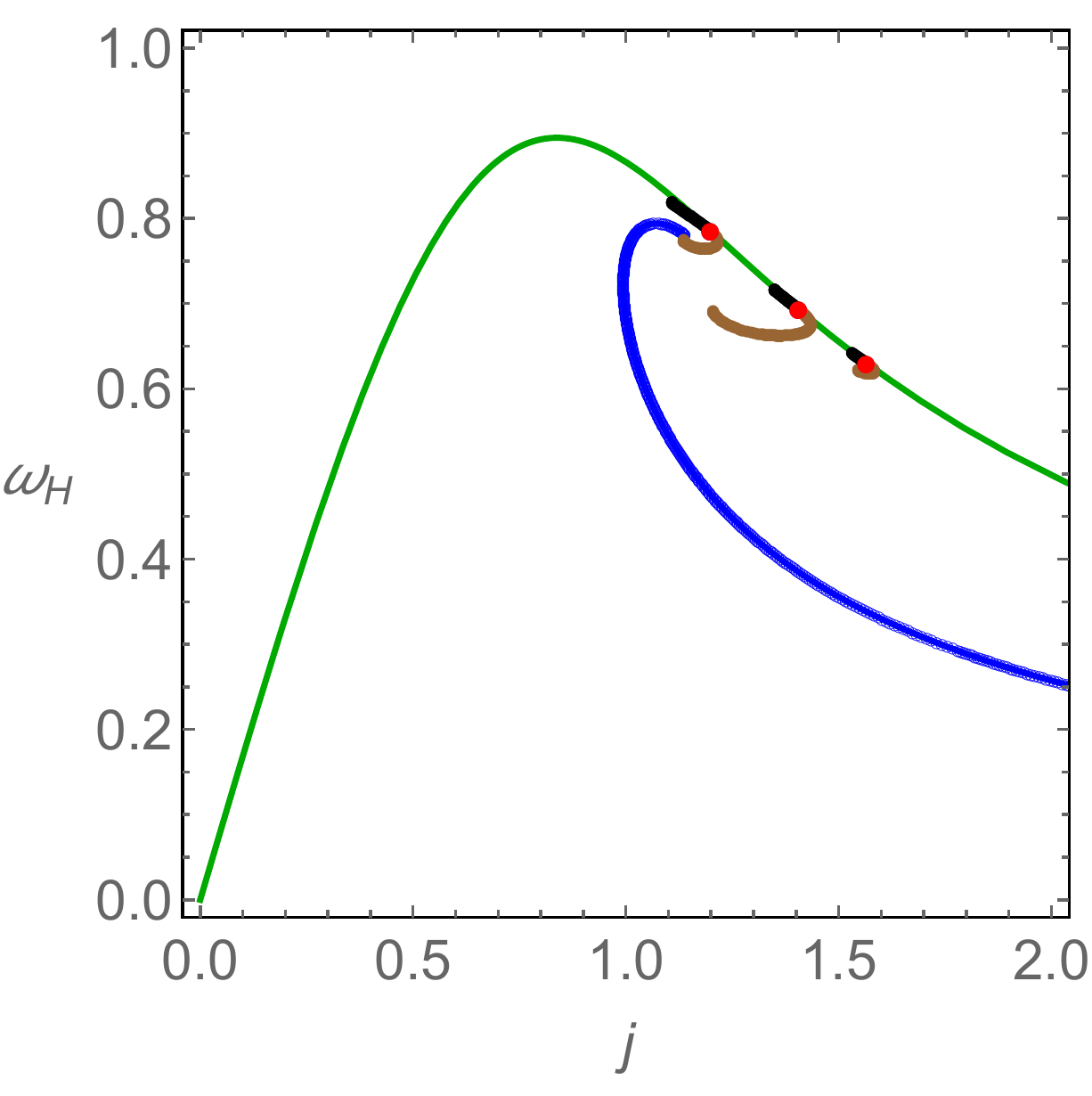}}\qquad
\subfigure[Zoom of (e)]{\includegraphics[scale=0.6]{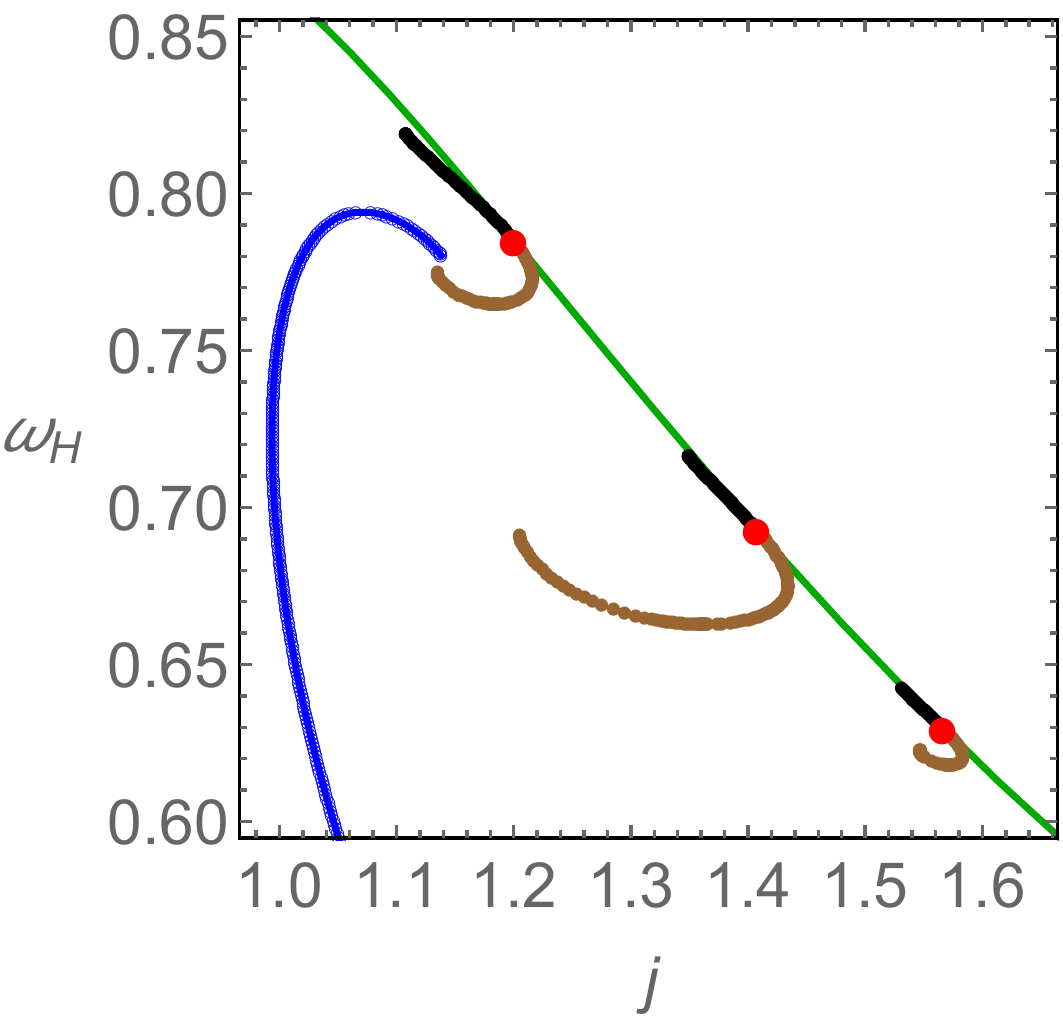}}
\caption{\small Thermodynamic quantities. Green: MP black hole. Brown: $(+)$-branch bumpy black holes. Black: $(-)$-branch bumpy black holes. Blue: black rings (from \cite{Dias:2014cia}). Red dots: branching points from the zero modes found in \cite{Dias:2010maa}. }
\label{thermoplots}
\end{figure}

\begin{figure}
  \centering
\subfigure[First bifurcation point]{\includegraphics[scale=0.36]{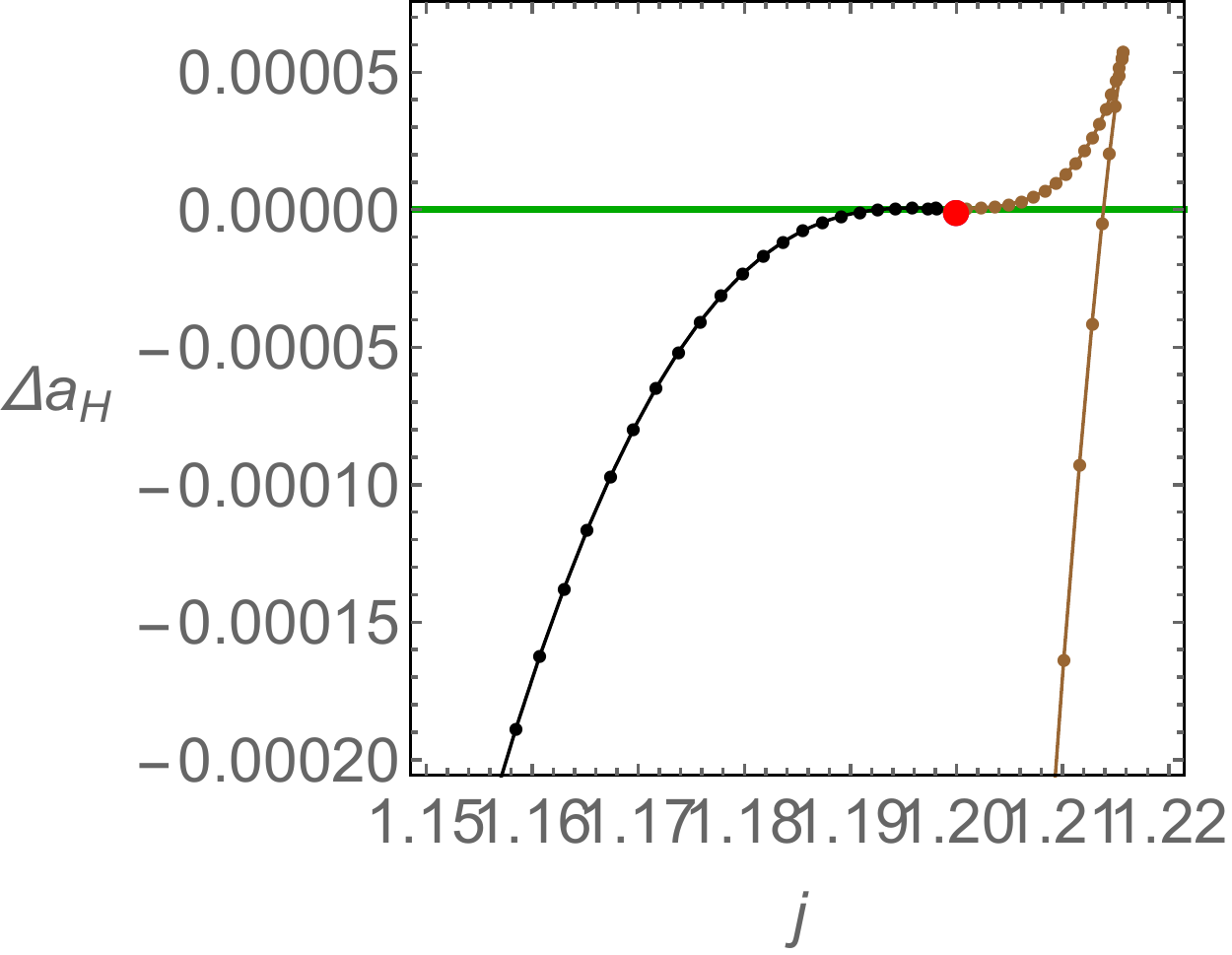}}\quad
\subfigure[Second bifurcation point]{\includegraphics[scale=0.36]{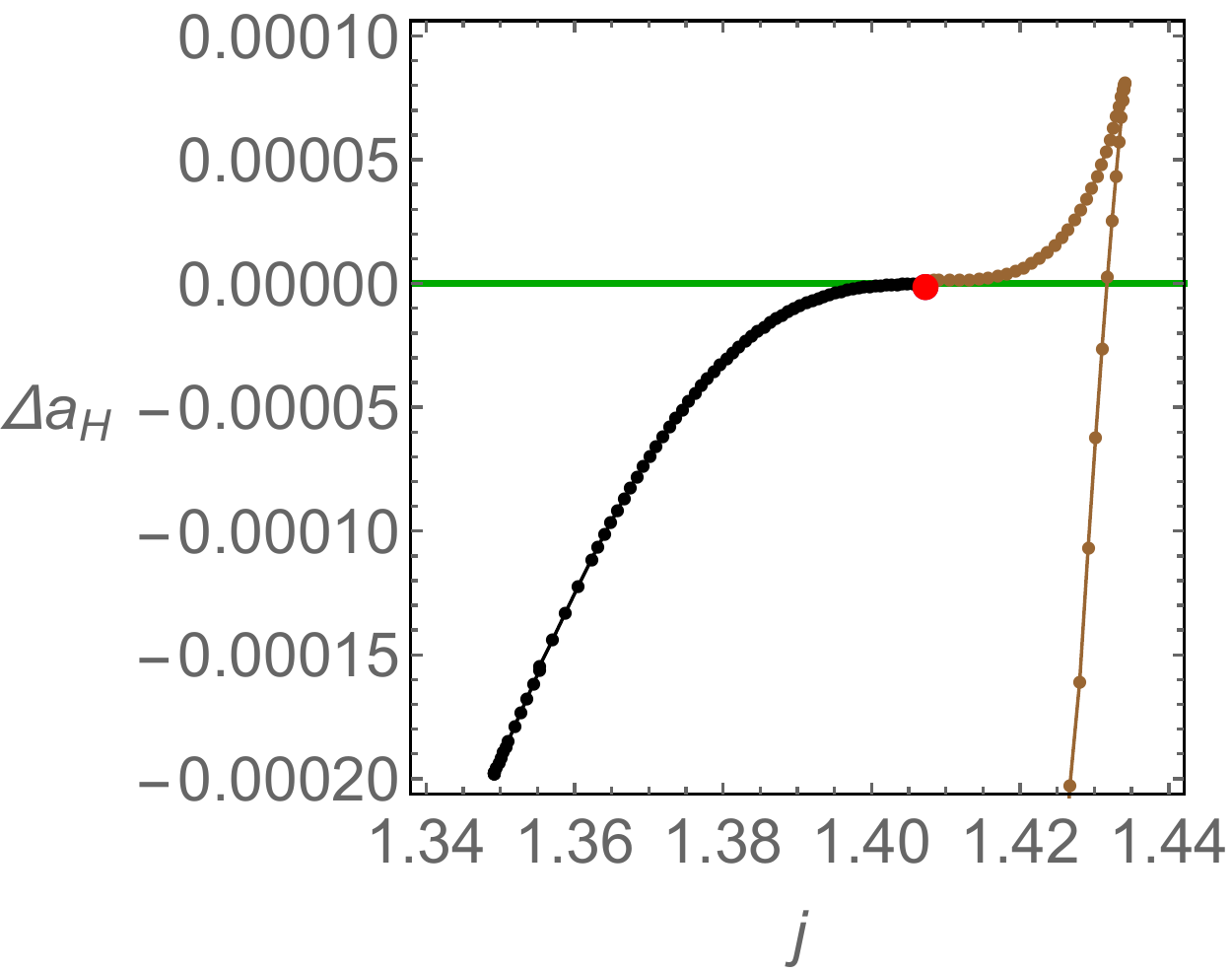}}\quad
\subfigure[Third bifurcation point]{\includegraphics[scale=0.36]{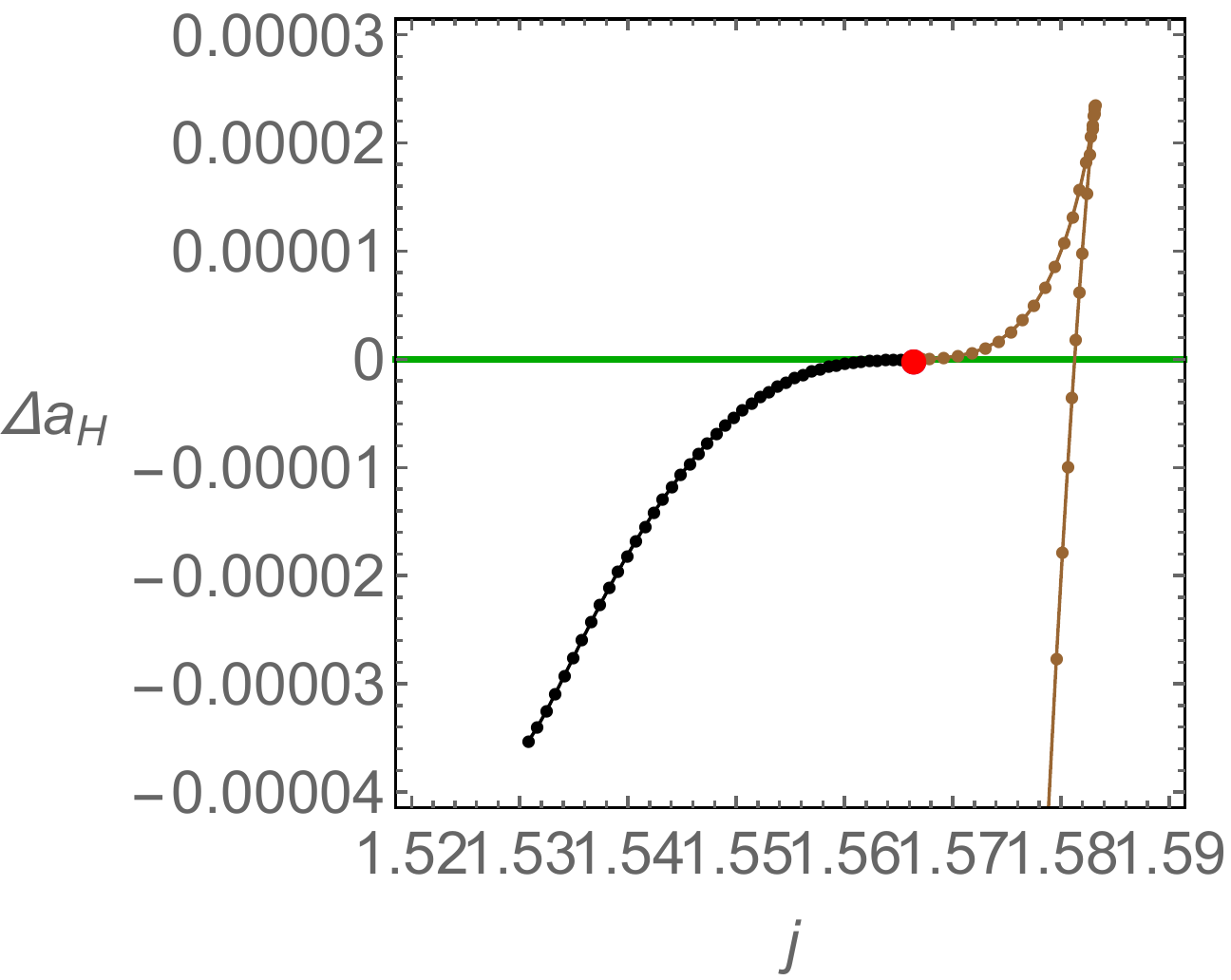}}
\caption{\small Area difference $\Delta a_{H}$ between bumpy and MP black holes vs.\ angular momentum $j$. Color coding as in fig.~\ref{thermoplots}.}
\label{deltaah}
\end{figure}

In fig.~\ref{thermoplots} we show the area, temperature and angular velocity as a function of the angular momentum for fixed mass. We can see the two different families of solutions branching off from each of the perturbative zero modes. Since in fig.~\ref{deltaah} it is difficult to distinguish the two branches in the $(j,a_{h})$ plane, we also show plots of the area difference between the bumpy black holes and the MP solutions for values of $j$ close to each branching point. The black ring phases obtained in \cite{Dias:2014cia} are also included in these plots, and it is apparent that the $(+)_1$ solutions tend to a merger point with the black rings. Although our results suggest that solution-trajectories inspiral close to this transition (which would lead to infinite discrete non-uniqueness of the kind found in \cite{Bhattacharyya:2010yg,Dias:2011tj,Gentle:2011kv}), our accuracy in this region is not enough to reach a definite conclusion.

Thermodynamic stability of these black holes in the grand-canonical ensemble is obtained when the specific heat at constant angular momentum $\mathcal{C}_j$ and the isothermal moment of inertia $\epsilon$ are both positive \cite{Monteiro:2009tc}
\begin{equation}\label{suscep}
\mathcal{C}_j=\left.\frac{dM}{d\mathcal{T}}\right|_J>0,\qquad
\epsilon=\left.\frac{dJ}{d\Omega_H}\right|_\mathcal{T}>0.
\end{equation}
Negative moments of inertia are possible for black holes since they are not rigid bodies. They can reduce their angular velocity while gaining angular momentum by spreading in the rotation plane. This is precisely what happens in ultra-spinning MP black holes. In this case it impossible for the black hole to remain in equilibrium with a co-rotating radiation reservoir. 

\begin{figure}
\centering
\subfigure[$M$ vs. $\mathcal{T}$ for fixed $J$. The slope is the specific heat at constant angular momentum $\mathcal{C}_J$]{\includegraphics[scale=0.65]{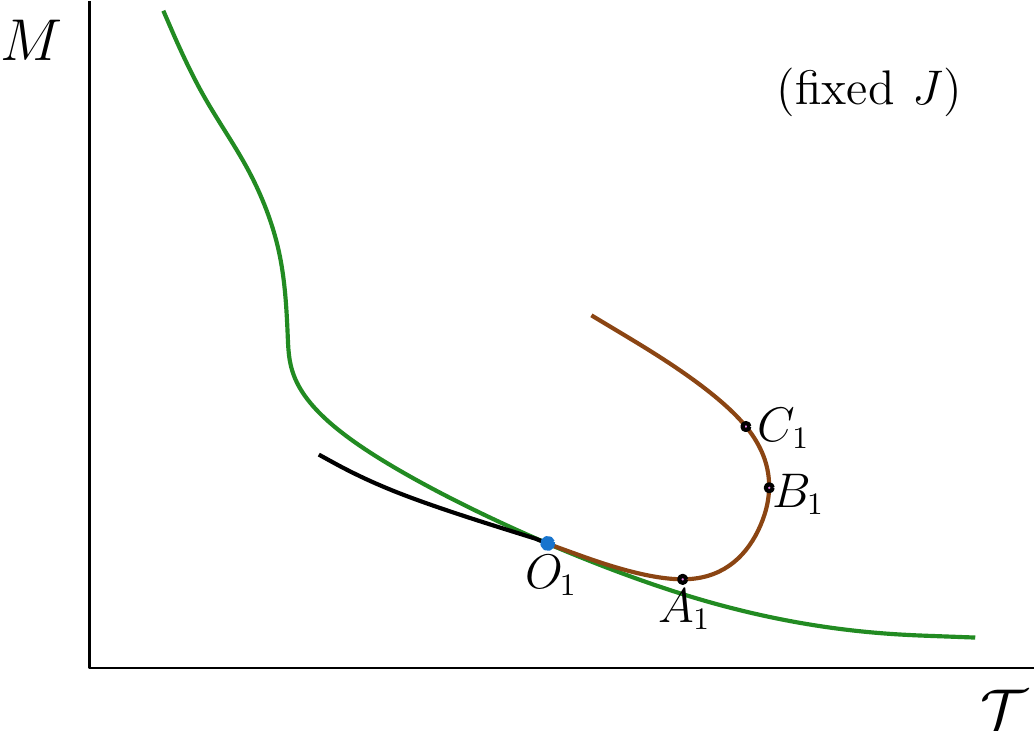}}\qquad
\subfigure[$J$ vs. $\Omega_H$ for fixed $\mathcal{T}$. The slope is the isothermal moment of inertia $\epsilon$]{\includegraphics[scale=0.65]{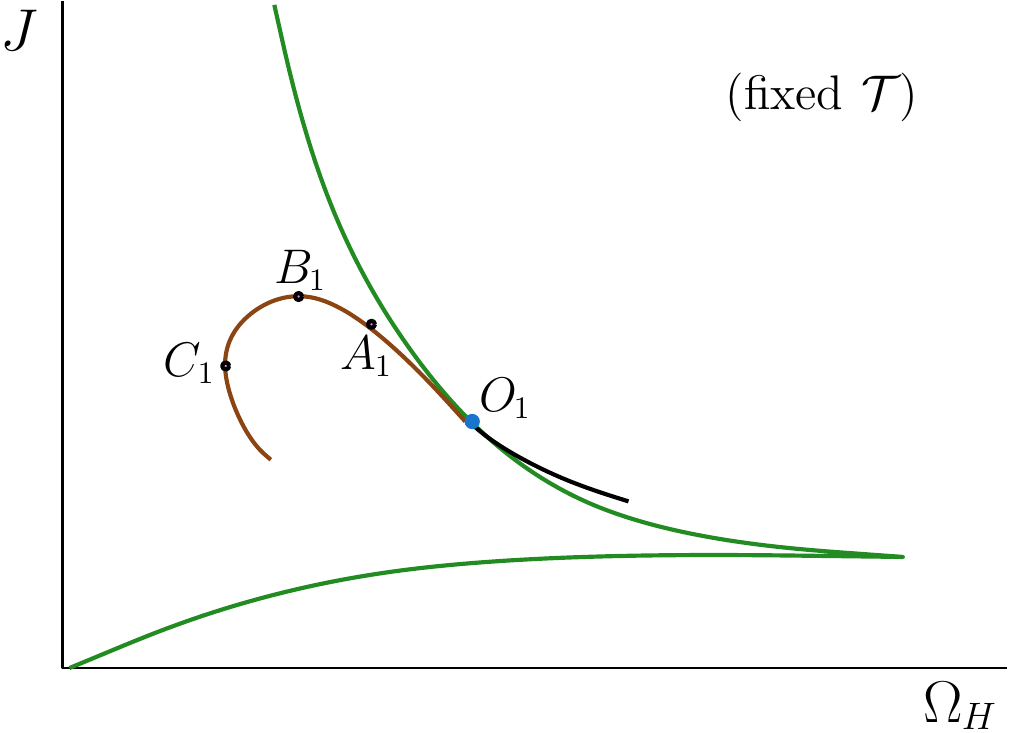}}
\caption{\small Sketch of phases in the $(\mathcal{T},M)$ and $(\Omega_H,J)$ planes (color coding  as in fig.~\ref{thermoplots}).  We only show the branches $(\pm)_1$, but $(\pm)_{2,3}$ have the same behavior. The branch $(+)_1$ (brown) has two negative modes from $O_1$ to $C_1$ and three from $C_1$ onwards, while the branch $(-)_1$ (black) always has three negative modes. Color coding as in fig.~\ref{thermoplots}.}
\label{cartoon}
\end{figure}

The specific heat and moment of inertia  can be read off from the slopes of the solution curves in the $(\mathcal{T},M)$ and $(\Omega_H,J)$ planes. The details of the plots for actual solutions are difficult to distinguish, so instead in fig.~\ref{cartoon} we present sketches of them that capture their qualitative features. 

In addition, we have also studied the spectrum of the Lichnerowicz operator, since its negative eigenvalues are directly related to the negative modes of the quasi-Euclidean action. We have checked that the number of negative eigenvalues coincides with the expectations from thermodynamic stability. In particular, along the MP family of solutions in the direction of increasing $j$, initially the solutions have one negative mode that corresponds to negative specific heat (it is the MP extension of the Euclidean Schwarzschild negative mode), and acquire a second one at the cusp in the $(\Omega_H,J)$ plane where the moment of inertia first becomes negative. This is also the minimum of the temperature (see fig.~\ref{thermoplots}) which signals the entrance into the ultraspinning regime, and which coincides with the change of sign of $\epsilon$. At higher $j$ one encounters further zero modes that become negative ones. These are not associated to new thermodynamic instabilities, instead they are `overtones' of Gregory-Laflamme-type negative modes.

The thermodynamic stability and negative modes along $(+)$ branches are more complicated, as there are several points where the susceptibilities \eqref{suscep} change sign. Here we explain it for $(+)_1$ solutions (higher $(+)$ branches exhibit the same qualitative behavior), referring to fig.~\ref{cartoon}:
\paragraph{From $O_1$ to $A_1$:}Point  $O_1$ is the bifurcation from the MP branch of solutions. The new branch bifurcates with higher area, hence the MP solution is expected to be less stable, and indeed it acquires an extra negative mode, while the bumpy solution keeps the number of negative modes present in the MP solutions just before $O_1$. Both the specific heat and the moment of inertia are negative in this segment; accordingly, the Lichnerowicz operator on solutions from $O_1$ to $A_1$ has two negative eigenvalues.
\paragraph{From $A_1$ to $B_1$:}The point $A_1$ at which $\mathcal{C}_j$ changes sign from negative to positive passing through zero  corresponds to the cusp in the $(j,a_h)$ plane in fig.~\ref{thermoplots}, where the branch beyond $A_1$ has lower area and $\mathcal{C}_j$ remains positive until $B_1$. But there is no qualitative change in the spectrum of the Lichnerowicz operator at $A_1$.
%
We interpret the two negative eigenvalues present here as due to the negative $\epsilon$ and to the fact that there exists another solution with higher area for the same $j$. Observe that a given negative mode does not strictly correspond to just one instability.
\paragraph{From $B_1$ to $C_1$:}At $B_1$ the sign of $\mathcal{C}_j$ changes from positive to negative and the sign of $\epsilon$ from negative to positive. Like before, the number of negative modes is preserved and the spectrum of the Lichnerowicz operator does not signal these changes in thermodynamic susceptibilities.
\paragraph{From $C_1$:}At $C_1$ the sign of  $\epsilon$ changes from positive to negative going through infinity. The Lichnerowicz operator acquires a third negative mode.

We see that the $(+)$-branch solutions are always thermodynamically unstable, since either  $\epsilon$ or $\mathcal{C}_J$ or both are negative. The solutions are likely dynamically unstable to bar-mode perturbations, like MP black holes are at even lower values of $j$.

\bigskip

Regarding the $(-)$ branches, they all have negative $\mathcal{C}_j$ and $\epsilon$. In addition they come out of the bifurcation with less area than the MP  black holes. As expected from the arguments above, the Lichnerowicz operator on these solution has three negative eigenvalues. We also expect them to be dynamically unstable to bar-mode deformations.

\section{Acknowledgements}

We are very grateful to \'Oscar Dias, Jorge Santos and Benson Way for very useful discussions and for generously supplying the numerical data for black rings used in fig.~\ref{thermoplots}. This work started during the COST Short Term Scientific Mission COST-STSM-MP1210-15750. MM thanks DAMTP, Cambridge, for kind hospitality during her visit. Part of this work was
done during the workshop ```Holographic vistas on Gravity and Strings'' YITP-T-14-1
at the Yukawa Institute for Theoretical Physics, Kyoto University, whose kind hospitality
we all acknowledge. MM is also grateful to the String Theory Group at NTU (Taiwan) for warm hospitality. RE and MM are supported by MEC FPA2010-20807-C02-02, FPA2013-46570-C2-2-P, AGAUR
2009-SGR-168 and CPAN CSD2007-00042 Consolider-Ingenio 2010. PF is supported by the European Research Council grant no.\ ERC-2011-StG279363-HiDGR and also by the Stephen Hawking Advanced Research Fellowship from Centre for Theoretical Cosmology, University of Cambridge. MM is supported by an FI Fellowship of AGAUR, Generalitat de Catalunya, 2013FI\_B 00840. 

\appendix

\section{Numerics}
\label{sec:numerics}
In this appendix we explain the details of our numerical construction of the bumpy black holes. 

Plugging the ansatz \eqref{ansatz} and the reference metric \eqref{refmetric} into the Einstein-DeTurck equations gives a system of partial differential equations that we solve numerically.

\begin{figure}
\centering
\includegraphics[scale=0.5]{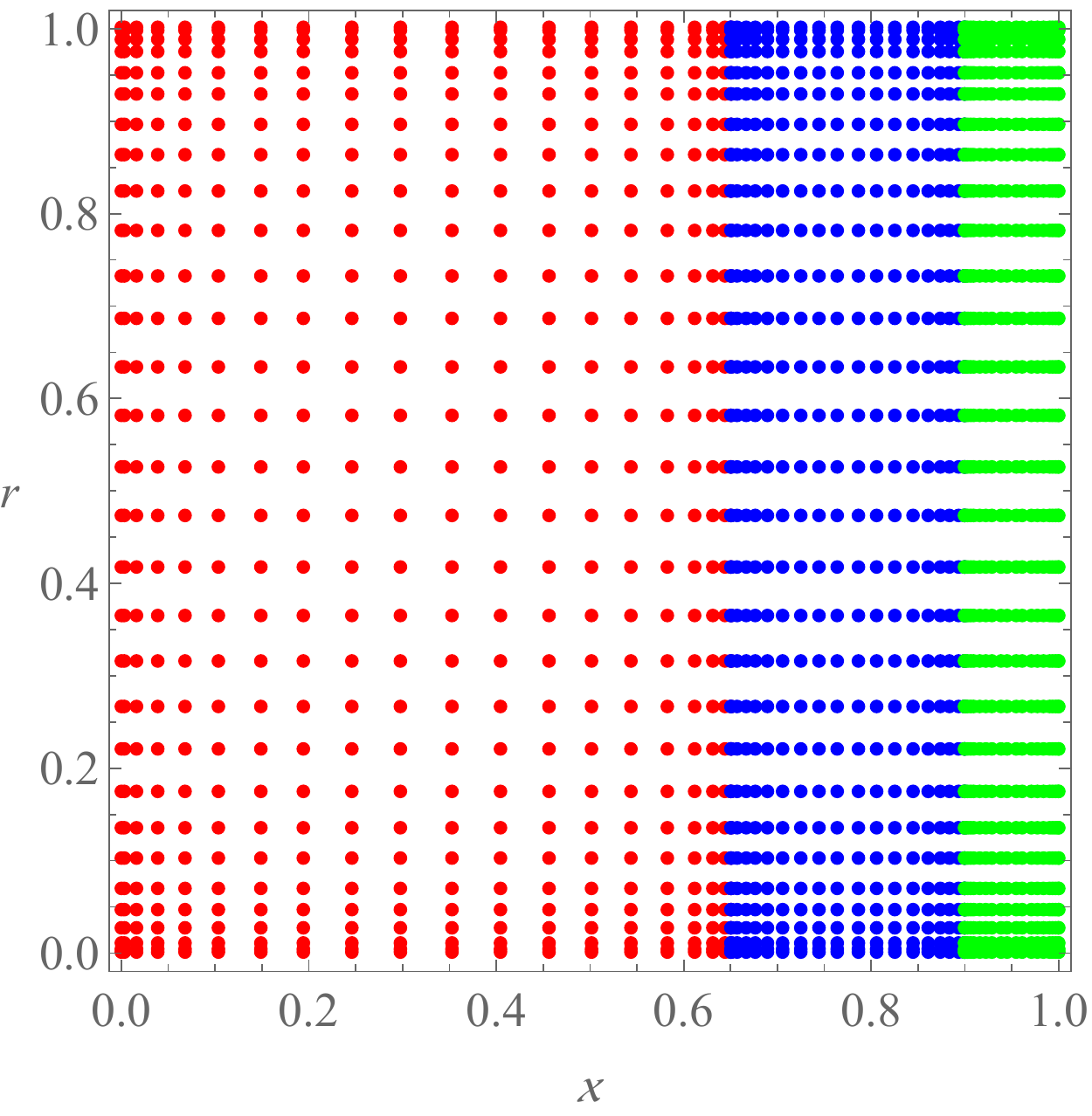}
\caption{\small Grid used for some solutions. Each of the three patches (red, blue, green) has 30 points in the $r$ direction and 20 in the $x$ direction.}
\label{domain}
\end{figure}

First we discretize the system using Chebyshev points. We need more resolution in the angular $x$ coordinate than in the radial $r$ one, so we use conforming patches, see fig. \ref{domain} for an example. This is computationally cheaper than having one bigger grid and gives us the flexibility of increasing the resolution just where it is necessary. This type of patches coincide along one line of points (no overlapping regions), in the present situation they coincided along a line of constant $x$. We used 2 to 5 patches depending on various factors. Higher zero modes have more bumps (the $Q$'s vary more along $x$) and we need more resolution. Close to transitions the functions become singular and therefore we need to concentrate more points in a specific part of the domain. We impose continuity of the functions and the first derivatives as boundary conditions between the patches.

Once discretized, we solve the system by an iterative Newton-Raphson method. Since this method needs a seed, we first solve the linearized problem, find the eigenvectors that correspond to the zero modes and use them (ref. metric perturbed with the eigenvector) as a seed for the first solution in each of the branches. Once we have solved the nonlinear problem, we move along the branch by using the previous solution as a seed and by changing the value of $k$ in the background metric. We keep $r_0=1$ in all the solutions.

For the standard branches that connect the MP black hole with the black ring, black Saturn and black diring, we begin by increasing the temperature. At some point the branches reach a maximum of the temperature and in order to go past it we keep $k$ fixed and vary $r_+$ instead. The solutions close to this maximum are tricky to obtain 
because the Lichnerowicz operator has a near zero mode,  but once we pass it the following solutions are easily obtained by lowering the temperature (decreasing $k$ with fixed $r_+$). The other type of branches do not have any extrema of the temperature and to obtain them we always decrease $k$.

As for the resolution used, we began with two patches of $20\times20$ in the branches $(+)_{1,2}$ (heading towards the black ring and black saturn) and with four patches of $30\times20$ for the $(+)_{3}$ branch (heading towards the diring); we began with similar resolutions for the other branches. 
In order to know when to increase the resolution we estimated the numerical error in the physical quantities and if it was greater than a few percent we decided that more resolution was needed. We have also checked that our numerical solutions converge to the continuum limit according to our discretization scheme.

\end{document}